\documentclass[pra,aps,superscriptaddress,twocolumn,floatfix,notitlepage,longbibliography]{revtex4-2}

\usepackage{amsmath}
\usepackage{amsfonts}
\usepackage{amssymb}
\usepackage{bm}
\usepackage{mathrsfs}
\usepackage{graphicx}
\usepackage{color}
\usepackage{xcolor}
\usepackage{xspace}
\usepackage{stmaryrd}
\usepackage{laussy}
\usepackage[sort&compress]{natbib}

\usepackage{esvect}

\usepackage[colorlinks, linkcolor=magenta, citecolor=magenta,
urlcolor=blue]{hyperref}
\newcommand{\av}[1]{\langle  #1 \rangle}
\newcommand{\tr}{\mathrm{Tr}}

\begin{document} 
\title{Polariton BECs\\Theory and Concepts}

\author{Fabrice P.~Laussy}
\email{fabrice.laussy@gmail.com}
\affiliation{Instituto de Ciencia de Materiales de Madrid ICMM-CSIC, 28049 Madrid, Spain}

\date{\today}

\begin{abstract}
Polaritons are a superposition of light and matter, that
  combine Strong Interferences (of light) with Weak Interactions (of
  excitons), making them WISI (Weakly-Interacting,
  Strongly-Interfering) particles. Their condensation is the main
  highlight of a field which occupies a unique position at the
  intersection of quantum optics, condensed matter physics and
  nonlinear dynamics of driven, dissipative systems.  This chapter
  surveys selected theoretical concepts of polariton condensates'
  formation, coherence and dynamics, with an emphasis on their
  distinctions from their atomic counterparts and on points of ongoing
  controversy. We argue that linear and non-interacting effects are
  undervalued in polariton physics, and that a significant part of the
  phenomenology---including bosonic correlations and coherence
  buildup---can often be understood without invoking strong
  interactions or genuine quantum effects.
\end{abstract}

\maketitle
\tableofcontents
\section{Introduction}

The term ``polariton'' was introduced by Hopfield~\cite{hopfield58a}
who studied the problem of propagation of light in crystals as part of
his PhD thesis, following a suggestion of his thesis adviser
Overhauser, who, however, did not participate in this research.
Hopfield reminisces that the polariton ``\emph{was invented to resolve
  the paradoxical situation [\dots] with the radiative lifetime of an
  exciton in a crystal, where the conflict was within theory
  itself. Naive theory yielded either zero or infinity depending on
  how it was applied, neither of which seemed to make sense.}''  This
problem was also solved by Pekar~\cite{pekar57a} and
Agranovitch~\cite{agranovich59a}, but the terminology of Hopfield
prevailed
. After graduation, Hopfield changed topic to neural networks, for
which he was awarded the 2024 Nobel Prize of Physics.  He attributes
the success of his solid-state publication ``\emph{to the existence of
  lasers, the polariton condensate, and modern
  photonics}''~\cite{hopfield14a}.  This text overviews some concepts
and theory of polariton condensates and the role they play in a wealth
of topics, including lasers and modern photonics. One of the latest
applications of polariton condensates is for neural
networks~\cite{ballarini20a,sedov25a,opala25a,gan25a}. Instead of the
usual consensual review-style presentation, we take here the approach
of contrasting various interpretations of what polariton condensation
is about, including whether this qualifies as BEC or as a laser,
whether this is a quantum or nonlinear classical phenomenon, or how
theories of greatly-varying sophistication, manage to describe the
dynamics of polariton condensation, in particular regarding the role
of interactions and nonlinearities. In this way, we hope to give a
breadth of what makes polariton condensation a still very active,
controversial and largely unresolved topic, which is a direct
consequence of how fundamental and versatile the problem remains as a
whole. This is also what makes it particularly rich and interesting.

\section{The Polariton quasi-particle}

\subsection{The discovery of the polariton}

The polariton is a quasi-particle in solids, very much like the
better-known phonon, which arises from the quantization of sound, as
is introduced in the undergrad physics
curriculum~\cite{kittel_book04a}. The specificity of polaritons is to
involve two other quasi-particles: the exciton, and the photon.  The
exciton itself---the hydrogen-like binding of a electron with the hole
it left in the valence band---is already an extremely complex
composite quasi-particle~\cite{knox_book63a}. It solves elegantly the
problem of the fundamental atomic excitation in a crystal, by
delocalizing it over all the atoms as a propagating quantum (whence
the name ``exciton'') whose translational motion preserves that of the
system.  Adding to that the question of the recombination into a
photon---or the absorption of a photon into an exciton---brought
Hopfield to their coexistence as a combined exciton-photon compound:
the polariton. In a quantum picture, both the bare exciton and the
bare photon fade into new eigenstates: half-light/half-matter new
quasi-particles, which Rabi oscillate the one into the other or
coexist as a quantum superposition, depending on the quantum state
they are prepared into.  This became a general concept that now
requires to specify the type of excitation to which the photon
couples. This could be to lattice vibrations (phonon-polaritons,
actually the first reported type~\cite{huang51a}) or a wealth of other
quasiparticles (plasmon-polaritons, magnon-polaritons, etc.)  The most
important type of polaritons remains the \emph{exciton-polariton},
since excitons in crystals are the fundamental material excitations
into which photons get converted~\cite{laussy24a}. This most important
type is, however, not precisely that of Hopfield, although he did
consider excitons.  But he also considered propagation in a bulk (3D)
material, where the polariton does not decay and thus remains
unobservable, except at the boundary of the solid, where the exciton
itself ceases to exist, so the quasi-particle has to carry on as a
photon only, producing luminescence. The observation of such photons
is thus an indirect---surface mediated---picture of the 3D polariton.
In a 2D system (such as a quantum well), the 2D exciton couples to 3D
photons and the excitation continuously leaks out in a
Wigner-Weisskopf fashion, failing to form a
polariton~\cite{toyozawa59a}. This can be remedied by restoring the
same dimensionality for photons by embedding the quantum well in a
microcavity, but nobody had this idea.  Claude Weisbuch discovered
such 2D-polaritons accidentally~\cite{weisbuch92a} while working on
VCSELs during a sabbatical in Arakawa's group in Tokyo. His aim was to
increase luminescence but he observed instead an unsuspected
anticrossing: ``\emph{The absence of any single peak no matter how
  hard I looked, I then decided that this was a true physical effect,
  and had to decide which}''~\cite{weisbuch05b}.  Ironically, Weisbuch
was one of the few expert experimentalists on Hopfield-type (bulk)
polaritons, having reported fifteen years earlier the first resonant
polariton fluorescence in Gallium Arsenide~\cite{weisbuch77a}. In
Tokyo, however, he chose to interpret the phenomenon as vacuum-field
Rabi splitting and even to oppose this picture to
exciton-polaritons. He did so to draw closer connections with the
cavity QED precedent set by atoms~\cite{raizen89a}. Although the
parallel is sound, it overlooks the propagation of polaritons. This
was quickly appreciated and restored under the denomination of
``cavity polaritons'' a few months later in a summer school (in
Cargese) organized by Weisbuch, where the new discovery was amply and
hotly debated~\cite{weisbuch_book95a}. The experimentally-discovered
cavity-polaritons turned out to be much more relevant than their
theoretically predicted bulk precursors, thanks to the reduced
dimensionality. In this case, the excitation is constantly held at the
surface and thus becomes directly accessible, furthermore with a
one-to-one mapping between the angle of emission and the wavevector of
propagation, providing a direct imaging of the distribution.  In the
bulk, in contrast, since polaritons remain out of reach until they
touch the surface, where they admix all wavevectors, the
reconstruction of their internal dynamics is hopeless. From now on, we
will call simply ``polariton'' this much more interesting Weisbuch
cavity-exciton-polariton, the quasi-particle that endows its atomic
counterpart with propagation.  The polariton discovery illustrates the
importance of experiments, even at the most basic and fundamental
level, in a field which cannot be defined as chiefly experimental or
theoretical. Nevertheless, this Chapter will focus on the theoretical
ideas and on their conceptual interpretations, with the guiding theme
that polaritons are great physical objects to which nothing seems out
of reach, although it is difficult to always agree on what it is that
they exactly do. Yet, condensates is what they do best, and are often
at the heart of the most important and interesting phenomena.  The
next Chapter by Ardizzone and Sanvitto addresses experimental
realizations.

\subsection{One-particle picture}

\begin{figure*}[t]
  \centering
  \includegraphics[width=\linewidth]{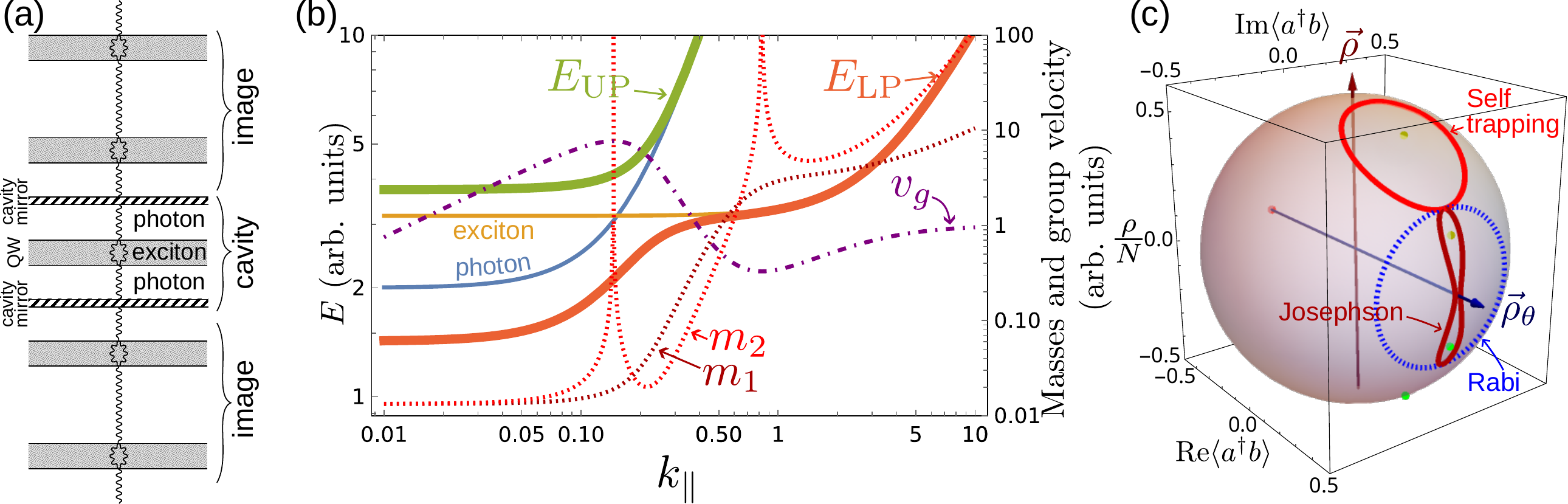}
  \caption[One-particle polariton picture]{ \textbf{The one-particle
      polariton picture.} (a) Sketch of a microcavity as two mirrors
    sandwiching a quantum well, respectively trapping photons and
    excitons, whose confinement brings in the strong-coupling
    regime. By the method of images, the cavity mirrors can be removed
    and the confined polariton becomes propagating again, as
    originally envisioned by Hopfield in bulk materials. (b) The
    lower~$E_\mathrm{LP}$ and upper~$E_\mathrm{UP}$ polariton
    dispersions, arising from the strong coupling of the exciton and
    photon parabolic dispersions with largely differing masses, so
    that the exciton appears flat over a broad range of momenta. The
    strong anharmonicity of the polariton dispersions leads to a
    plethora of sui generis properties, including two divergences of
    the diffusive mass~$m_2$ ($m_1$ is the inertial mass) and a
    non-monotonous group velocity~$v_g$. Solid lines have axes on the
    left and broken lines on the right. (c) Polariton dynamics on the
    Bloch sphere: Rabi dynamics is a circle on the sphere (dashed
    blue) circling around the polariton axis~$\vv*{\rho}{\theta}$,
    whose relation to the laboratory axis~$\vv\rho$ determines the
    oscillating dynamics. Here at resonance~$\Delta=0$,
    $\vv*{\rho}{\theta}\perp\vv{\rho}$. With interactions, new fixed
    points appear which distort the circle and can drive it into the
    Josephson (purple) regime, with possibility of self-trapping (red)
    around one fixed point only.}
  \label{fig:ThuJan1121128PMCET2026}
\end{figure*}

In symbols and reduced to its simplest expression (one mode), a
single polariton is a quantum superposition of a cavity photon~$\ket{1,0}$
with a material excitation (exciton)~$\ket{0,1}$. Those bare states
couple to each other to yield two dressed states (in atomic
language), known as the L(ower) and U(pper) Polaritons (in
solid-state language):
\begin{equation}
  \label{eq:Sun24Nov180218CET2024}
  \ket{\psi}_\mathrm{LP/UP}\equiv\alpha_\mathrm{LP/UP}(t)\ket{1,0}+\beta_\mathrm{LP/UP}(t)\ket{0,1}\,.
\end{equation}
In this popular and almost unanimous picture, the polariton is a
quantum mélange of light and matter. As per the axioms of quantum
theory, the coefficients~$\alpha$ and~$\beta$ are probability
amplitudes, whose modulus square corresponds to the probability that,
upon measurement, the polariton will decide (collapse) whether it is a
photon or an exciton.  Hopfield's bulk propagation and Weisbuch's
frozen vacuum Rabi splitting can be reconciled by Yamamoto~\emph{et
  al.}~\cite{yamamoto93a}'s unfolding of the light-matter interaction
by the method of images, as reproduced in
Fig.~\ref{fig:ThuJan1121128PMCET2026}(a). The cavity mirror frees the
confined photon and corrects Weisbuch's dictum that ``\emph{there is
  no exciton polariton like in the 3D semiconductor
  case}''~\cite{weisbuch92a}. In both cases (bulk and cavity), the
polariton decays when the photon hits the end of the chain: the
surface in bulk, or the imperfect reflection of the cavity.  The
coefficients~$\alpha$ and~$\beta$ in
Eq.~(\ref{eq:Sun24Nov180218CET2024}) are two complex numbers, and can
thus be mapped to the Bloch sphere.  As such, a polariton could be
understood as a qubit. This defines a first notion of ``polaritonic
quantum state'', in the sense of which diametrically-opposed points on
the Bloch sphere define the lower and upper polaritons.  For
real-valued coefficients, the one-polariton quantum states read
$\ket{\mathrm{LP}}=\kket{1,0}$ and $\ket{\mathrm{UP}}=\kket{0,1}$ with
\begin{subequations}
   \label{eq:ThuJan1113111AMCET2026}
  \begin{align}
   \kket{1,0}&=\cos\theta\ket{1,0}-\sin\theta\ket{0,1}\,,\\
   \kket{0,1}&=\sin\theta\ket{1,0}+\cos\theta\ket{0,1}\,,
  \end{align}
\end{subequations}
%
where~$\theta$---the ``mixing angle''---can be time dependent. If it
is not, this describes stationary states---or eigenstates---of the
system, which can also vary as a function of other parameters, typically
the wavevector. Indeed, since polaritons are two-dimensional systems,
and are formed from particles with different masses, those can only
have the same energy (be in resonance) for a single momentum.  This
does not have to be the~$k=0$ (non-propagating) case, but depends on
the photon-exciton detuning.  The exciton dispersion is the standard
one for a massive particle:
\begin{equation}
  \label{eq:ThuJan1124039PMCET2026}
  E_\mathrm{ex}(\mathbf{k})=E_\mathrm{ex}(0)+{\hbar^2k^2\over2m_\mathrm{ex}}
\end{equation}
where $E_\mathrm{ex}(0)$ is the rest energy of the bound electron-hole
state, and $m_\mathrm{ex}$ is the effective mass of the exciton, which
combines the individual effective masses of both the electron and hole
into a measure of how the bound electron-hole pair responds to
external forces within the crystal lattice, and determines key
excitonic properties such as the Bohr radius and the
binding energy.  On the range of wavevectors relevant for light-matter
coupling, this mass is so large that the dispersion can be considered
flat.  For the cavity photon, due to confinement, its dispersion is
roughly parabolic since, starting from $\omega_\mathrm{C}=kc/n$,
with~$c$ the speed of light and $n$ the cavity refractive index,
freezing the perpendicular component
of~$k=\sqrt{k_\parallel^2+k_\perp^2}$ to the fundamental mode of a
cavity with effective length~$L_\mathrm{C}$, i.e.,
$k_\perp=\pi/L_\mathrm{C}$, the first order
expansion~$\sqrt{1+\varepsilon}\approx 1+\varepsilon/2$ yields the
cavity-photon dispersion
$E_\mathrm{C}(\mathbf{k})\equiv\hbar\omega_\mathrm{C}$ as:

%
%
\begin{equation}
  \label{eq:SunAug24023426PMCEST2025}
  E_\mathrm{C}(\mathbf{k})=E_\mathrm{C}(0)+{\hbar^2k^2\over2m_\mathrm{C}}
\end{equation}
where~$m_\mathrm{C}\equiv {\hbar\pi n/(cL_\mathrm{C})}$ is the
effective cavity-photon mass,
$E_\mathrm{C}(0)\equiv\pi\hbar c/(nL_\mathrm{C})$ is the rest mass,
and we now write~$k$ only for the in-plane~$k_\parallel$ component of
the 2D particle.  The factor~$n^2$ between the vacuum rest-mass energy
$m_\mathrm{C}c^2=E_\mathrm{C}(0)n^2$ and the medium rest-mass
energy~$E_\mathrm{C}(0)$ is due to the dielectric medium slowing down
the group velocity, with higher refractive index resulting in stronger
polarization response, in turn increasing the stored energy
density. The resulting cavity-photon mass is about four orders of
magnitude smaller than the typical exciton mass, so although photons
are no longer massless particles, they now make excitons look like
particles of infinite mass.  Including a momentum-independent
exciton-photon coupling~$\Omega_\mathrm{R}$ (R is for Rabi), we arrive
to the one-particle polariton Hamiltonian
\begin{equation}
  \label{eq:ThuJan1012203PMCET2026}
  H=\sum_\mathbf{k} E_\mathrm{C}(\mathbf{k})\ud{a_\mathbf{k}}a_\mathbf{k}
+E_\mathrm{ex}(\mathbf{k})\ud{b_\mathbf{k}}b_\mathbf{k}+{\hbar\Omega_\mathrm{R}\over2}(\ud{a_\mathbf{k}}b_\mathbf{k}+a_\mathbf{k}\ud{b_\mathbf{k}})\,,
\end{equation}
which---since the photon~$a_\mathbf{k}$ and exciton~$b_\mathbf{k}$ are
boson operators---can be diagonalized into uncoupled,
freely-propagating particles with bosonic operators~$p_\mathbf{k}$
and~$q_\mathbf{k}$:
\begin{equation}
  \label{eq:ThuJan1012555PMCET2026}
  H=\sum_\mathbf{k}E_\mathrm{LP}(\mathbf{k})\ud{p_\mathbf{k}}p_\mathbf{k}
  +E_\mathrm{UP}(\mathbf{k})\ud{q_\mathbf{k}}q_\mathbf{k}\,.
\end{equation}
The lower~$p$ and upper~$q$ polariton field operators are still
bosons, with dispersions:
\begin{multline}
  \label{eq:ThuJan1013440PMCET2026}
  E_{\mathrm{UP}\atop\mathrm{LP}}(\mathbf{k})\equiv{E_\mathrm{C}(\mathbf{k})+E_\mathrm{ex}(\mathbf{k})\over 2}\pm\\{1\over2}\sqrt{\big(E_\mathrm{C}(\mathbf{k})-E_\mathrm{ex}(\mathbf{k})\big)^2+\hbar^2\Omega_\mathrm{R}^2}.
\end{multline}
These one-particle properties are the foundation for polariton physics
and contain many of the critical aspects of their description, even
for condensation, although one does not make condensates with only one
polariton. A selling point is their small effective mass, since the
critical temperature for condensation scales inversely with the mass
of the particles. This was already an argument for excitons and in the
case of polaritons, their admixture with photons further halves this
at resonance, and still smaller masses are possible for more
photon-like polaritons.  Such properties result in the peculiar
polariton dispersion, which rules the free-particles dynamics.  The
case of a positive detuning---where the cavity has higher energy than
the quantum well at normal incidence---is shown in
Fig.~\ref{fig:ThuJan1121128PMCET2026}(b), in which case the maximum
superposition of light and matter~$\alpha=\beta=1/\sqrt2$ occurs for a
finite~$k$.  The polariton dispersion was first measured, by
angle-resolved spectroscopy, by Houdr\'e \emph{et
  al.}~\cite{houdre94b} and its first detailed theoretical analysis
was performed by Savona \emph{et al.}~\cite{savona95a,savona95b} among
others~\cite{pau95a}.  The single-mode picture of
Eq.~(\ref{eq:Sun24Nov180218CET2024}) describes plane waves, with
well-defined momenta. Real systems are localized in space and thus
involve various wavevectors. Since the exciton and photon contents
vary with momentum, the light-matter
superposition~(\ref{eq:Sun24Nov180218CET2024}) is more complex for
spatially-extended fields, as they admix polaritonic quantum states,
even for single particles. We will come back to this.

\subsection{Many-body polaritons}
\label{sec:SatJan3103901AMCET2026}

Condensation is a many-body phenomenon, so we must upgrade our
previous picture to multiple particles. If a polariton is best
understood as a superposition of a photon with one exciton, how are we
to understand a collection of several polaritons? This brings us to
the second concept of quantum state, this time not the ``polaritonic
quantum state'' previously introduced as the exciton/photon
composition of a single polariton, i.e., a point on the Bloch sphere,
but in the quantum-optical sense of what type of pure or mixed density
matrix one has for the collection of particles.

\begin{figure*}
  \centering
  \includegraphics[width=\linewidth]{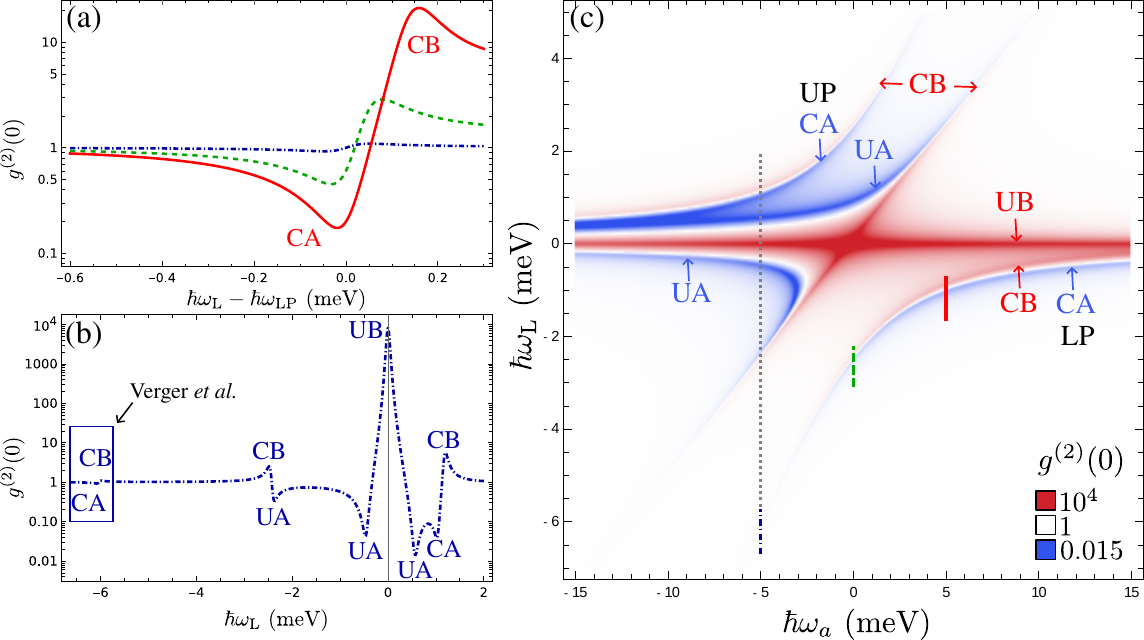}
  \caption{\textbf{The two-particle polariton picture.} (a) Polariton
    blockade with $g^{(2)}(0)<1$ as a function of the laser detuning
    from the lower polariton branch, for three exciton-photon
    detuning, with the dashed green line at resonance, solid-red at
    positive cavity detuning (exciton-like LP) and dash-dotted black
    at negative cavity detuning (photon-like LP). (b) Same as~(a) for
    the negative detuning case and over an enlarged spanning window of
    laser energy (now shown relative to the exciton), to also include
    the UP branch, with panel~(a) identified in the box
    $\approx-6$meV. Considerably stronger correlations are accessible
    around other resonances of the system, most of them not visible in
    photoluminescence as being off-branch. (c) Full two-photon
    correlation picture revealed by the CUBA
    (Conventional/Unconventional Bunching/Antibunching) formalism with
    $g^{(2)}(0)$ of the emitted photons shown as a function of the
    exciton-photon detuning~$\Delta$ and driving laser
    frequency~$\omega_\mathrm{L}$. Red shows bunching~$g^{(2)}>1$,
    blue shows antibunching~$g^{(2)}<1$ and white shows no
    correlation~$g^{(2)}=1$. Besides the familiar LP and UP branches
    new resonances of the weakly-interacting polariton condensate are
    revealed in this two-photon observable.  Conventional features
    arise from the dressed states while unconventional ones arise from
    interferences. Antibunching is much stronger around the UP branch,
    not only because it is exciton-like, but also and mainly for its
    proximity to unconventional antibunching. The cuts shown in
    Panels~(a) and~(b) are identified by the vertical lines.  All
    figures adapted from Ref.~\cite{zubizarretacasalengua20a} for the
    parameters of Ref.~\cite{verger06a}.}
    \label{fig:SunJan4111733PMCET2026}
\end{figure*}

The simplest many-polariton state is the one with, say, $n$ lower
polaritons, which can be calculated by iterated operations of the
lower-polariton creation operator~$\ud{p}$, which is expressed in
terms of the photon~$\ud{a}$ and exciton~$\ud{b}$ creation operators
as the unitary relation that links the one-particle
states~(\ref{eq:ThuJan1113111AMCET2026}), i.e.,
$p\equiv\cos(\theta)a-\sin(\theta)b$, in which case, $n$ polaritons
indeed result in a strongly entangled superposition of $k$ multiphoton
and~$(n-k)$ multi-exciton states~$\ket{k,n-k}$:
\begin{equation}
  \label{eq:MonSep15092237PMCEST2025}
  \kket{n,0}=\sum_{k=0}^n\sqrt{\binom{n}{k}}\cos^k\theta\sin^{n-k}\theta\ket{k,n-k}\,.
\end{equation}
This most natural and straightforward passage from the one-particle
polariton---as the widely proclaimed light-matter quantum
superposition---to the purest version of condensation in the form of
\emph{all} particles piling up the ones onto the others in a single
mode, and with a fixed number of particles, is however too crude a
picture. Below the condensation threshold, the expected many-polariton
quantum state should be described by a thermal state, whose effective
temperature~$\theta\equiv n_\mathrm{th}/(1+n_\mathrm{th})$ gives the
mean occupation~$n_\mathrm{th}$. For, say, the ground state with
only~$\mathbf{k}=\mathbf{0}$ lower polaritons:
\begin{equation}
  \label{eq:SunJan11112228AMWET2026}
  \rho_\mathrm{th}=\sum_{n=0}^\infty{n_\mathrm{th}^n\over(1+n_\mathrm{th})^{1+n}}\kket{n,0}\bbra{n,0}\,.
\end{equation}
This is a mixed state, i.e., with no off-diagonal elements.  The
density matrix for the full system is then simply the product state (no
correlations) of the various modes,
$\rho=\bigotimes_{\mathbf k}\rho_{\mathbf{k},\mathrm{th}}$, with the
population for each mode given by the Bose-Einstein
distribution~$n_{\mathbf{k},\mathrm{th}}=1/(e^{(E_\mathrm{LP}(\mathbf{k}) -
  \mu)/k_B T} - 1)$, again assuming only the lower branch.

Polaritons are typically contemplated through their photon projection,
in which case one must trace out the exciton component in
Eq.~(\ref{eq:MonSep15092237PMCEST2025}) or
Eq.~(\ref{eq:SunJan11112228AMWET2026}), as the case maybe, which
results in a binomial distribution for the Fock-state condensation and
another (photonic) thermal state for thermal polaritons. The binomial
distribution provides the expected picture of condensation in the
limit of large particles, where the distribution then approaches the
awaited Poisson distribution for a coherent state, but one would still
need to worry about aspects of large particle numbers, related to
interactions and the excitonic structure, to which we will come
back. For now, we remain within the linear aspect of many-body
polariton physics, so at low densities.

The field dynamics of polaritons is typically captured with a
one-particle wavefunction~$\psi(\mathbf{r},t)$. When this describes,
not a single particle as in the previous section, but a condensate,
then all particles are assumed to be in the same state, and one can
use a one-particle wavefunction for the many-body problem. This is the
approach initially taken by Gross~\cite{gross61a} and
Pitaevskii~\cite{pitaevskii61a} to provide their eponymous equation
where the many-body physics is translated into a nonlinear
term. Instead, we can consider the single-particle field not as a
mean-field but as an operator which, applied on the vacuum, creates a
particle at the position~$\mathbf{r}$, as
$\hat\psi^{\dagger}(\mathbf{r})\ket{\mathrm{vac}}=\ket{\mathbf{r}}$.
This yields the density operator
$\hat{n}^{(1)}(\mathbf{r}) \equiv \hat\psi^\dagger(\mathbf{r})
\hat\psi (\mathbf{r})$, whose average on the many-body quantum state
yields the one-particle reduced density matrix, which is equivalent to
the modulus square of the wavefunction~$|\psi(\mathbf{r},t)|^2$,
with~$t$ coming from the quantum state.  We could upgrade the density
matrix to its exciton-photon degrees of freedom to describe both
branches, but we focus on lower polaritons only, for simplicity.  The
object of interest for a multi-particle theory of lower-polariton
condensation is the many-body density matrix
$\hat\rho\equiv\sum_{\mathbf{m},\mathbf{n}}\alpha_{\mathbf{n}\atop\mathbf{m}}\ket{\mathbf{n}}
\bra{\mathbf{m}}$ where~$\mathbf{n}$, $\mathbf{m}$ are vectors of
integers that specify how many modes in a suitable basis are occupied
in the occupation-number (second quantization) formalism.  For
instance,
$\ket{\mathbf{n}}=\ket{n_{\mathbf
    0},n_\mathbf{k_1},\cdots,n_\mathbf{k_m}}$ refers to the state
with~$n_{\mathbf{k}_i}$ particles in a plane wave of
wavevector~$\mathbf{k}_i$ ($n_{\mathbf{0}}$ are particles in the
ground state). Instead of such a standard plane-waves representation,
Zubizarreta Casalengua and the author have used the Laguerre-Gaussian
basis~$\phi_n$ to describe topologically interesting states such as
vortices~\cite{arXiv_casalengua24a} (which, incidentally, were also
the concern of both Gross and Pitaevskii). In this case, the creation
operator can be expanded
as~$\hat\Psi^{\dagger}(\mathbf{r})\equiv\sum_m \phi_m^{*}(\mathbf{r})
a_m^{\dag}$ where~$a_m$ is the bosonic operator for the corresponding
LG mode. This provides us with a formalism to describe quantum-optical
states of spatially extended fields. The one-body densiy matrix for
instance, is recovered as
$\rho^{(1)}(\mathbf{r}) = \sum_{p,q} \av{a_p^{\dagger} a_q}
\phi_p^*(\mathbf{r}) \phi_q(\mathbf{r})$ where
$\av{a_p^{\dagger} a_q}\equiv\tr{\hat\rho a_p^{\dagger}
  a_q}=\sum_{\mathbf{m},\mathbf{n}}\alpha_{\mathbf{n}\atop\mathbf{m}}\bra{\mathbf{m}}\ud{a_p}a_q\ket{\mathbf{n}}$. Similarly---and
more importantly since this brings us to the many-particle aspect of
the problem---the reduced two-polariton density
matrix~$\rho^{(2)} (\mathbf{r},\mathbf{r}') = \langle{:}\hat{n}^{(1)}
(\mathbf{r}) \hat{n}^{(1)}(\mathbf{r}'){:}\rangle$ is obtained
as~\cite{arXiv_casalengua24a}
$ \rho^{(2)} (\mathbf{r},\mathbf{r}') = \sum_{p,p',q,q'}
\av{a_p^{\dagger} a_{p'}^{\dagger} a_{q'} a_q} \phi_p^* (\mathbf{r})
\phi_{p'}^* (\mathbf{r}')\phi_q(\mathbf{r})\phi_{q'}(\mathbf{r}')$.
The same could be done for any number of particles. We can now
characterize quantum features from multi-polariton observables, for
instance, the probability for particles to be detected closely
together, as is expected from bosons as a result of Bose
stimulation. Taking the case of two LG$_0^{\pm1}$ vortices with
different circulation, their two-body density matrix, obtained as just
described, is
$\rho^{(2)}_\mathrm{B} (\mathbf{r},\mathbf{r}') = |\phi_\la
(\mathbf{r}) \phi_\ra (\mathbf{r}')+\phi_\ra (\mathbf{r}) \phi_\la
(\mathbf{r}')|^2$. This indeed maximizes the probability to find both
particles together (twice as much as if they would be uncorrelated),
but also similarly maximizes their probability to be found on opposite
sides of the donut, i.e., it maximizes their alignment, not their
proximity. If one had two fermions instead, the wavefunction would be
antisymmetric
$\rho^{(2)}_\mathrm{F} (\mathbf{r},\mathbf{r}') = |\phi_\la
(\mathbf{r}) \phi_\ra (\mathbf{r}')-\phi_\ra (\mathbf{r}) \phi_\la
(\mathbf{r}')|^2$, thereby maximizing the probability for the
particles to be found at 90$^\circ$ angles the one from the other,
which is closer on average than half the bosonic case which get
diametrically opposed. In this way, and counter-intuitively, if one
seeks to maximize the distance between two particles, one should opt
for bosons. They will often be found very close, but also very far, in
a bimodal distribution of distances, while fermions tend to remain
within similar distances. This is for exactly two bosons, or
structureless polaritons. For thermal states---the case below
condensation threshold---the two-polariton density matrix reads
$\rho^{(2)}_\mathrm{th}(\mathbf{r},\mathbf{r}') =
\sum_{p,p'\in\{\la,\ra\}} {n}_{\mathrm{th},p} {n}_{\mathrm{th},p'}
\big(|\phi_p(\mathbf{r})|^2 |\phi_{p'}(\mathbf{r}')|^2 +\phi^*_p
(\mathbf{r}) \phi_p (\mathbf{r}') \phi^*_{p'} (\mathbf{r}') \phi_{p'}
(\mathbf{r}) \big)$ with the same type of correlations than for
exactly two particles, but weaker. For instance, while two bosons can
never be found at a right-angle on the donut from two Fock states, the
probability remains small but nonzero in the case of thermal
states. Importantly, in all those cases, the one-body density matrix
remains the same: a donut, with the vortex core at the center. From
one-particle observables, therefore, one cannot distinguish a Fock
states of two particles, from a thermal state or from a broad family
of other multi-particle quantum states. To make such distinctions, one
must consider at least two-body observables.  Condensates do not
reduce to how many particles are found in one mode of the system, but
in which many-body quantum state they are found to populate this
mode. This can only be resolved from quantum observables.

\subsection{Quantum vs Classical pictures of polaritons}
\label{sec:ThuJan1112943AMCET2026}

The description of polaritons as quantum objects---namely, the quantum
superposition of light and matter as expressed by
Eq.~(\ref{eq:Sun24Nov180218CET2024})---has been a temptation since the
early days and has remained the default understanding. Romuald Houdré,
who took up the polariton problem immediately after Weisbuch,
recollects: ``\emph{we had in mind all the very exciting concepts
  developed in quantum optics and we did not worry much about an old
  paper in atomic physics by Y. F. Zhu et al., arguing that even in
  atomic physics, this so-called vacuum field Rabi splitting did not
  require any quantum field theory or even quantum mechanics to be
  explained and that, probably quantum effects, if any, were to be
  searched in more subtle properties and
  experiments.}''~\cite{houdre05a} The ``old'' Zhu \emph{et al.}
paper is Ref.~\cite{zhu90a} and was not old at the time. This is part
of the polariton folklore to attribute their---indeed
unique---characteristics to quantum properties. It is also naturally
assumed that those extend straightforwardly to the many-body case too,
and thus one eagerly speaks of quantum fluids of
light~\cite{carusotto13a}. However, much of the polariton
phenomenology can also be explained without any quantum field theory
or even quantum mechanics. The quantum picture merely happens to be
conceptually simpler to describe coupled oscillators and their
eigenstates. But in reality, an accurate description of polaritons in
typical scenarios is not in terms of Fock states, with the popular
picture of exactly one particle at all time, Rabi-oscillating between
the photon and exciton fields between which it periodically lingers as
a quantum superposition. This indeed gives us the first equation of
this text. It could be the case, but this requires the corresponding
quantum preparation, which was implemented by the Sanvitto
group~\cite{suarezforero20a} but no sooner than 2020, and by exciting
the microcavity with a single-photon source, following an ad hoc
theoretical proposal to create genuine quantum polaritons: bring the
quantumness from outside~\cite{lopezcarreno15a}. In this case, one
excites a true quantum of the polariton field, in essence,
transferring a quantum (the photon) into the cavity, where it couples
to the exciton and fulfills the scenario of
Eq.~(\ref{eq:Sun24Nov180218CET2024}). Depending on the details of the
photon (its frequency, momentum, wavepacket profile, etc.), one can
prepare different initial conditions.  In a usual scenario, where the
drive itself is classical (a laser), one gets instead the coherent
excitation of the vacuum through the displacement
operator~$D(\alpha)\equiv\exp\left(\alpha a^\dagger - \alpha^*
  a\right)$, which creates the coherent state of Glauber
\begin{equation}
  \label{eq:MonSep15092709PMCEST2025}
  \ket{\alpha}=\exp(-|\alpha|^2/2)\sum_{n=0}^\infty{\alpha^n\over\sqrt{n!}}\ket{n}\,,
\end{equation}
where~$\ket{n}$ here is~$n$ photons in the presence of the exciton
vacuum---assuming that the laser couples to the cavity photons as is
the usual assumption---so we could write~$\ket{n,0}$ in the notation
of the rhs of Eq.~(\ref{eq:MonSep15092237PMCEST2025}). There is no
entanglement, since $\ket{n,0}=\ket{n}\otimes\ket{0}$ and the exciton
vacuum factors out. To show that no entanglement ever forms in
such a scenario, let us assume instead a coherent state not of
photons, but of lower polaritons, i.e., let us
replace~$\ket{n}\equiv\ket{n,0}$ in
Eq.~(\ref{eq:MonSep15092709PMCEST2025}) by~$\kket{n,0}$, the strongly
entangled multi-polariton Fock
state~(\ref{eq:MonSep15092237PMCEST2025}).  Since the claim to
quantumness of the polariton resides in its quantum superposition of
light and matter, let us rewrite the coherent lower-polariton state
$\kket{\alpha}_\mathrm{LP}$ in the photon/exciton basis.
Inserting Eq.~(\ref{eq:MonSep15092237PMCEST2025}) in
Eq.~(\ref{eq:MonSep15092709PMCEST2025}), one finds that binomial
identities disentangle the fields:
\begin{equation}
  \label{eq:MonSep15092023PMCEST2025}
  \kket{\alpha}_\mathrm{LP}=\ket{\alpha\cos\theta}_a\ket{\alpha\sin\theta}_b
\end{equation}
i.e., a polariton coherent state is a separable product of coherent
states of photons and excitons. There is no entanglement and no
superpositions anymore. This remains true with dynamics, since the
initial state~$\ket{\alpha_0}\ket{\beta_0}$ in the photon-exciton
basis evolves~as:
\begin{equation}
  \label{eq:MonSep15110149PMCEST2025}
  \ket{\alpha_0\cos(gt)-i\beta_0\sin(gt)}
  \ket{-i\alpha_0\sin(gt)+\beta_0\cos(gt)}
\end{equation}
which is simply the physics of two coupled harmonic oscillator in ket
notations: what a downgrade from the entangled quantum vacuum! But
what a refreshing exercise to visualize polaritons as what they really
are at the classical level: harmonic oscillators. The particular
case~$\alpha_0=\mp\beta_0$ turns
Eq.~(\ref{eq:MonSep15110149PMCEST2025})
into~$\ket{\alpha_0 e^{\pm igt}}\ket{\mp\alpha_0 e^{\pm igt}}$ which,
back in the polariton basis, corresponds to, respectively, the lower
polariton coherent
state~$\skket{-\sqrt{2}\alpha_0e^{igt}}_\mathrm{LP}\kket{0}_{\mathrm{UP}}$
with the vacuum upper-polariton field (which we previously didn't
write),
and~$\kket{0}_\mathrm{LP}\skket{\sqrt{2}\alpha_0e^{-igt}}_{\mathrm{UP}}$.
Lower polaritons should thus be visualized as out-of-phase
exciton-photon fields going against the rotating frame, hence lowering
their energy, while upper polaritons are in-phase fields adding their
Rabi rotation to the rotating frame, hence increasing their
energy. Rabi oscillations are obtained for non-eigenstate initial
conditions, such as~$\alpha_0\neq 0$ and~$\beta_0=0$ (injecting
photons at~$t=0$) in which case
Rabi-oscillations~$\ket{\alpha_0\cos(gt)}\ket{-i\alpha_0\sin(gt)}$
oscillate now in quadrature (perpendicularly) and radially, as opposed
to, for polaritons, staying on the circle.  This disentangling of
light and matter through the coherent state is a general
phenomenon. In the many-body picture of spatially extended polariton
condensates of Section~\ref{sec:SatJan3103901AMCET2026}, for instance,
while thermal (below threshold) or Fock (quantum) states exhibit
correlations, the coherent states display no
correlations~\cite{arXiv_casalengua24a}: particles are distributed
independently when sampled from a coherent state, turning their
wavefunction into a true one-particle density probability, with no
withdrawn information for higher particle-number observables,
justifying the mean-field treatment. This correspondence between the
classical and quantum (correlated) cases is complete through the
mathematical identity of the~$\alpha(t)$ and~$\beta(t)$ coefficients
in both the polariton Fock (quantum) state
Eq.~(\ref{eq:Sun24Nov180218CET2024}) and the (classical) normal-mode
coupling of two oscillators as a product of coherent
states~$\ket{\alpha(t)}\ket{\beta(t)}$: they obey the same equations
and thus undergo the same dynamics in time. But their interpretation
is completely different: in the former case, they are Hopfield
coefficients, i.e., quantum probability amplitudes, 
while in the latter case, they are classical-field amplitudes in the
sense of Maxwell's electrodynamics, whose modulus square provides
continuously-varying intensities for both fields to be jointly present
in the corresponding amounts. This mapping between the two cases can
lead to confusion.  Their enthusiastic association has been made for
the polariton condensate dynamics by Alexey Kavokin \emph{et al.} who
introduced the idea of a polariton qubit~\cite{demirchyan14a}. In its
initial and simplest formulation, however, the more likely relevant
theoretical framework is that of a classical analogue, or so-called
cebit~\cite{spreeuw01a}. 
Through such a formalism, one can map the dynamics of \emph{classical}
polaritons on the Bloch sphere---the emblem of the qubit---and
accurately describe in this way the Rabi dynamics observed in
classical experiments, such as those performed by Dominici \emph{et
  al.}~\cite{dominici14a}'s ultrafast (with fs timescale) coherent
control of polaritonic quantum states (i.e., which type of
polaritons), albeit all in coherent states (i.e., classical
states). 
Important and largely overlooked aspects of Dominici \emph{et al.}'s
work are that i) the coupling of the external laser is to both the
exciton and photon fields, not exclusively to the cavity as is usually
assumed, ii) the upper polaritons have an intrinsic lifetime in
addition to the photon and exciton lifetimes and iii) there is a
residual incoherent pumping of the excitons. Those specificities are
usually neglected in treatments of the polariton dynamics.

\section{Polariton Coherence}
\subsection{BEC vs Lasing}

Central to the polariton dynamics is their bosonic character, and
central to Bose dynamics is Bose stimulation.  Two fundamental
phenomena directly follow: i) Bose--Einstein condensation and ii)
Lasing. These are deeply intermingled yet separate phenomena.  One
could say that the former is due to stimulated \emph{scattering}
(particles end-up preferably in the ground state) while the latter is
due to stimulated \emph{emission} (a photon can clone another
one). Another popular distinction is that the former holds at
equilibrium while the latter is sustained
out-of-equilibrium. 
Bose--Einstein Condensation first arose as a theoretical concept, from
no less than Einstein himself~\cite{einstein25a} in the continuation
of his extension of Bose's statistics to massive
particles~\cite{einstein24a}, where he observed that ``\emph{das
  entartete Gas von dem Gas der mechanischen Statistik in analoger
  Weise ab, wie die Strahlung gemäß dem Planckschen Gesetz von der
  Strahlung gemäß dem Wienschen Gesetz}'' (``the gas of Bosons differs
from the gas of mechanical statistics in an analogous manner as the
radiation according to Planck's law differs from the radiation
according to Wien's law.'') Notably, this insight hinges on the
hypothesis of ``\emph{einer weitgehenden formalen Verwandtschaft
  zwischen Strahlung und Gas}'' (``a far-reaching formal similarity
between radiation and gas''), which befits polaritons. Besides,
Einstein also theorized stimulated emission~\cite{einstein17a}. Since
he did not come up with the idea of lasing (this emerged with the
maser in the year Einstein would die~\cite{gordon55a}) and arrived at
the BEC from thermodynamical arguments, the problem of their
connections and relationships were left for those coming after him.
Fr\"ohlich was the first to consider the possibility of strongly
out-of-equilibrium coherence powered by the channeling of excess
energy into a single mode~\cite{frohlich68b}. Curiously, such
considerations arose in the context of living organisms, namely,
longitudinal electronic modes of a cell's membrane~\cite{frohlich68a},
which he related to mechanisms well beyond the scope of physics, such
as cell-division or cancer. The principle is, however, that of
Einstein's 1925 argument: in a steady state, pumping balances losses,
and if those arise from a redistribution of the energy between the
available modes---featuring Bose stimulation---there is a
threshold of pumping beyond which the particle distributions cannot
accommodate all the excitations, and the surplus must therefore build
up (condense) into a single-mode. Fr\"ohlich observed that this
``\emph{closely resembles the condensation phenomenon in a Bose gas}''
(and even makes a stronger association in the
title~\cite{frohlich68b}). He did not connect it to lasing, but modern
commentators do~\cite{zhang19d}. Although directly related, Fr\"ohlich
condensation did not gain traction in the polariton community, but it
did in other related systems, such as magnon
condensation~\cite{xu25a}.

The term LASER---initially an acronym substituting the earlier MASER
that was Amplifying Microwaves (as opposed to Light) by Stimulated
Emission of their Radiation---overlooks a crucial aspect of the
dynamics: the positive feedback, provided by the cavity. Without it,
merely amplification (as the name states) is obtained. Oscillations,
rather than Amplifications, are needed to yield gain and the eventual
bright and coherent emission, but then the acronym would not be so
catchy. In BEC, there is no such positive feedback, and references to
lasing for processes that rely on BEC for coherence buildup---as is
the case for polaritons---face an immediate identity crisis. A
``polariton laser'' arises from the spontaneous emission of photons
from a polariton condensate itself maintained by external pumping. It
is thus unclear if such a phenomenon involves a BEC, a laser, both or
none. Such questions have been and remain central to the study of
polariton condensates. They actually do not reduce to mere
terminology, although this has been, so far, the most expedient
solution to this conundrum.  BEC as a concept predates lasing by three
decades. Experimentally, the demonstrations went the other way around,
with the first laser built in 1960~\cite{maiman60a}, preceding the
first realization of a BEC in 1995~\cite{anderson95a,davis95a} by
three and a half decades. In parts, this is due to lasers being
needed for BEC, to provide laser-cooling, but beside this technical
necessity, no further connection was made at the time between the two
concepts.  This experimental achievement---even before it awarded a
Nobel prize to the group leaders (Wieman \& Cornell, and Ketterle, in
2001)---rekindled considerable interest in all fields dealing with
bosons. Among them, excitons in particular already had a rich
theoretical literature of their
condensation~\cite{blatt62a,keldysh68a,hanamura77a} and even several
claims of their realization~\cite{peyghambarian83a,lin93a,butov94a}.
In fact, before the atomic demonstration, excitons were widely
regarded as ideal candidates (better than atoms) to demonstrate
Einstein's new phase of matter: their small mass means a high critical
temperature, and their coupling to light allows to control the
particles density merely by varying the optical excitation, as well as
to track in real time their dynamics through their luminescence.  This
gives direct access to their distribution in any desired phase space,
such as their energy/momentum distribution.  The short radiative
lifetime however makes the phenomenon intrinsically out-of-equilibrium
and excitons pioneered the central problem of finite-lifetime particle
condensation: do they have time? Dark excitons---uncoupled to
light---offered an escape but then also lost the ability to be
monitored, making the role of light both a blessing and a curse.  Post
atomic-BEC, excitons have remained a controversial topic regarding
their condensation despite compelling
evidence~\cite{butov02a,snoke02a} and remarkable
phenomenology~\cite{butov02b,snoke02b}.  Polaritons, on the other
hand, by relying mainly on the same conceptual advantages, managed
even more compelling demonstrations. To some, the fact that adding the
photon helps ``markers of condensation'' suggests that light, not
matter, gets coherent and thus coherence comes from lasing (from
light), not from a BEC (from matter)~\cite{butov07a,butov12a}. Still,
the polariton being a superposition, the recognition of one attribute
or the other to one component or the other did not convince, and
polariton condensation quickly garnered a broad enough consensus and a
large follow-up, with the community tacitly agreeing to set the
milestone of their BEC with the 2006 report by Kasprzak \emph{et
  al.}~\cite{kasprzak06a} (a prior claim in 2002~\cite{deng02a}, which
does not feature the BE of C in its title, did not get that widespread
recognition, and an earlier claim was withdrawn
altogether~\cite{cao97a}).  Theoretically, the earliest proposal for
polariton condensation is a little-known paper by Yura and
Hanamura~\cite{yura94a}, two years after Weisbuch's discovery of 2D
polaritons, and also two years before the better known similar
proposal by \Imamoglu \emph{et al}~\cite{imamoglu96a}. In all cases,
polariton condensation and/or lasing are very much tied to their
underlying excitonic versions.  The discrimination between those two
cases was more pronounced in the early days. For instance, it was
central to Imamoglu \emph{et al.}'s seminal considerations, who depict
the polariton laser as an interpolation between a BEC of excitons
(long lifetime, reaching an equilibrium) and a photon laser (with
population inversion, strongly
out-of-equilibrium)~\cite{imamoglu96a}. Given the trendiness of BEC,
this encouraged them to avoid polaritons so as to stay as far as
possible from lasing, with an ad hoc ``\emph{configuration [that]
  ensures that the relevant quasi-particles are excitons rather than
  polaritons}''~\cite{imamoglu96b}.  Besides, or in addition to, their
connections with excitons, one of the major challenges for polariton
condensation has been to distinguish polariton condensation from
photon lasing, adding to the semantic issue the nature of which
particles play a role in the coherent phenomena which are,
undoubtedly, observed. At too-high densities of excitons, for
instance, polaritons are destroyed and coherence follows from a VCSEL
scenario of population inversion of an electron-hole plasma. Criteria
emerged to ensure that whatever mechanism was taking place,
polaritons---not photons in weak coupling---are the particles
involved, for instance by checking that the dispersion relation is the
characteristic polariton one and not that of the cavity photons.
Various names have been proposed to circumvent the BEC vs lasing
quandary, including \Imamoglu and Pau's ``\emph{boser}'' or Baumberg
\emph{et al.}'s ``\emph{plaser}''. None of those terms stuck and
people settled instead with euphemisms like a quasi-condensate or
out-of-equilibrium condensate, as this question remains unresolved to
this day.  Atoms found much less contentious grounds to articulate
this question, since in their case, both implementations are
interesting, whereas lasing is expected in one form or another for
quasiparticles made of---or bathing in---light~\cite{bjork91a}. Lasing
was, in fact, Weisbuch's initial motivation. The concept of an atom
laser~\cite{wiseman95a, spreeuw95a, holland96a, mewes97a}, on the
other hand, does not downplay condensation but opens a new
discipline~\cite{meystre_book01a}. Rather than a plight, this identity
crisis of polaritons must be welcomed as a unique window on the
fundamental question of how BEC and lasing relate to each other.

\subsection{Polariton interactions}

So far, our discussion has focused on non-interacting
polaritons. Their specificities, in this case, rely on their
light-matter constitution, which in turn results in i) a peculiar
dispersion relation as well as ii) a short lifetime, iii) a high
degree of coherence with strong interferences and iv) a small
effective mass. This is inherited from the photon component and is
enough to make polaritons unique objects on their own. There is an
additional highly-prized property which they possess, this time from
the excitonic component: interactions. There is an erroneous tendency
to value polariton interactions as the main if not only attribute
worthy of interest in polaritons, seeing them as strongly-interacting
photons. First, polariton interactions are strong for photons but not
so much for material particles, and one needs high densities to see
obvious manifestation of direct polariton-polariton
interactions. Besides, an interesting observation made by Keeling and
Berloff~\cite{keeling11a} is that too-strong interactions are actually
detrimental for condensation, as they would hamper the separation of
single-particle modes between highly occupied and practically empty
ones. All modes couple to the whole system in very-strongly
interacting systems.  A better qualification for polaritons is
therefore as weakly-interacting strongly-interfering particles.
Besides, even ``weak'' interactions bring many complications at the
high densities required to make interactions compelling, and make the
simple picture at best a convenient approximation with loose
quantitative links to the experimental reality~\cite{snoke23a}. In
particular, direct polariton-polariton interactions may not be
dominant as compared to other mechanisms that provide similar
phenomenology. One is phase-space filling: the shortage of available
states in momentum or real space, for creating additional excitons due
to Pauli blocking of the underlying electrons/holes. This limits the
population but also reduces the exciton oscillator strength, and thus
also reduces the light-matter coupling and the Rabi splitting.  This
results in an apparent blue-shift of the condensate for increasing
densities, which is instead being quenched as the system
saturates. Another important mechanism that can produce false
signatures of direct interactions is the exciton reservoir, which
consists of excitons (or exciton-like polaritons) of high momentum,
which are those most easily formed under incoherent pumping, and also
those with the longer lifetime, including infinite lifetime if they
remain outside of the light cone. Therefore, one can have a large
population of free carriers acting as a potential which can result in
strong effects, even resulting in effective attractive polariton
interactions~\cite{vishnevsky14a}. Such effective potential energy
shifts are also not self-interactions from the condensate, but look
very much like them. Strong potential effects can be more easily and
effectively obtained by confinement, for instance by etching the
cavity to confine the polaritons~\cite{reithmaier97a}.  Above the
condensation threshold, such reservoir-excitons can be efficiently
sucked into the condensate and thus deplete the reservoir, which in
turns reduces the blueshift of the condensate~\cite{estrecho19a}.  The
concept of polariton itself become endangered at high densities, since
excitons break down as a Mott transition into an electron-hole
plasma. This is estimated to occur at excitonic
densities~$n_\mathrm{ex}\approx 1/a_\mathrm{B}^2$, setting an upper
boundary.  Polariton interactions nevertheless play a central role in
various phenomena, particularly within the dominion of nonlinear
optics, e.g., involving optical switching or
bistability~\cite{baas04a,bajoni08d}. Those typically involve
``condensates'' too but the focus is more on classical nonlinear
optics.  One of the age-making experiments of polaritonics was the
report in 2000 by Savvidis \emph{et al.} of parametric amplification
and oscillations by resonant driving of the polariton
branch~\cite{savvidis00a}, whereby the authors observed huge
amplification of a weak probe at normal incidence when matching
energy-momentum conservation on the polariton branch by pumping a
``magic angle''.  Recognizing the role of stimulated scattering, this
demonstrated the bosonic character of polaritons and triggered an
intense activity taking advantage of this peculiar mix: external
control, Bose stimulation and strong nonlinearities. This can be used
to implement a broad family of devices~\cite{sanvitto16a}. A beautiful
and simple three-mode (signal, pump, idler) theory by Ciuti \emph{et
  al.}~\cite{ciuti00a} was quickly provided that confirmed the
mechanism of all-polariton huge bosonic amplification.  Such
interactions also play a central role in more fundamental aspects of
condensations or superfluidity, in particular through the concept of
spectrum of elementary excitations. The Savvidis \emph{et al.}
pioneering bosonic stimulation as well as the first theoretical
picture by Ciuti assumed a strong adherence to the polariton
dispersion, which provides in particular energy/momentum conditions of
the type~$E_\mathrm{LP}(0)+E_\mathrm{LP}(2k_p)=2\hbar\omega_p$, i.e.,
four-wave mixing: two pump photons with wavevector~$k_p$ scatter into
the~$k=0$ ground state and idler~$2k_p$. It was quickly realized that
under strong driving or, equivalently, for strong interactions,
renormalization of the branches leads to off-branch virtual
states. Under coherent excitation, this can be accounted for fully
classically, as was done among others but prominently by
Gippius~\cite{gippius04a}, although it can also be fruitfully
described in a quantum formalism. In absence of resonant driving, such
elementary excitation spectra get us closer to the atomic situation
where such renormalizations result from the intrinsic many-body
dynamics of the condensate and accounts for its properties, in
particular superfluidity. Manifestations of interactions in this
context include the evidence of a Bogoliubov-like linear spectrum of
excitation which renormalizes as
$\sqrt{E_\mathrm{LP}(\mathbf{k})(E_\mathrm{LP}(\mathbf{k})+2gn)}$ and
which has been reported in photoluminescence~\cite{utsunomiya08a}.
Besides, the scattered and correlated character of the quantum
depletion of the condensate yields negative bogoliubov branches, i.e.,
with lower energies than the condensate. In the case of polaritons,
and in contrast to atomic systems, dissipations and the presence of a
reservoir lead to a $3\times3$ matrix for the linearization of the
fluctuations, leading in turn to three branches, including a flat,
dispersionless branch, tied to the fluctuations in the
reservoir~\cite{byrnes12b}. Such negative branches have also been
reported experimentally from a variety of
techniques~(pump-probe~\cite{kohnle11a} or in the photoluminescence of
an expanding polariton condensate~\cite{pieczarka15a}).  Although we
will see that the role of interactions for condensation is not as
crucial as is often claimed, still, given both their importance and
popularity in, at least the discourse on polariton condensation, we
must address their theoretical grounds in some detail.  What is clear
and consensual is that this is a complicated problem, with little that
is firmly established.  Even without direct interactions,
excitons---and therefore polaritons---are complex multi-particle
objects due to the indistinguishability of their constituting
electrons and holes. The main issue can be formulated as follows: when
picturing two excitons, one imagines two pairs of one electron and a
hole, but since those particles are indistinguishable, which electron
(or hole) is in which exciton? This causes considerable, in fact
possibly fundamental difficulties regarding the exciton-exciton
scattering rate, which according to some, in particular Monique
Combescot~\cite{combescot01a,combescot02b,combescot05b,combescot07a},
might not even be a well defined concept. For practical purposes, the
bulk of the literature treats such exciton-exciton physics as an
effective interaction, although its origin is in the geometric
constraints on the wavefunctions.  This upgrades the Hamiltonian of
Eq.~(\ref{eq:ThuJan1012203PMCET2026}) with an exciton-exciton
interaction term which would be fairly standard for structureless
particles (here neglecting spin):
\begin{equation}
  \label{eq:MonJan5061854PMCET2026}
  H_{\mathrm{ex-ex}}=\sum_{\mathbf{k}_1,\mathbf{k}_2,\mathbf{q}}M_{\mathbf{k}_1,\mathbf{k}_2,\mathbf{q}}\ud{b_{\mathbf{k}_1+\mathbf{q}}}\ud{b_{\mathbf{k}_2-\mathbf{q}}}b_{\mathbf{k}_1}b_{\mathbf{k}_2}\,.
\end{equation}
The matrix element~$M$ was first calculated by Ciuti \emph{et
  al.}~\cite{ciuti98a}, relying on the standard Born approximation and
overlap integrals of the 1s wavefunction with
Eq.~(\ref{eq:MonJan5061854PMCET2026}), arriving to the conclusion that
interactions can be well described by short range two-body
interactions with a $s$-wave scattering length of the order of the
exciton Bohr radius~$a_\mathrm{B}$ and with, for low exchanged-momenta
$k_1,k_2,q\ll a_\mathrm{B}$,
\begin{equation}
  \label{eq:MonJan5062743PMCET2026}
  M\approx 6{E_\mathrm{B}a_\mathrm{B}^2/S}  
\end{equation}
with~$S$ a normalizing sample area, which cancels when involving
wavefunctions.  This calculation attributes the dominant term of the
exciton–exciton interaction to the electron-hole exchange term, at
least for same-spin excitons, which is the main result (and title) of
the paper.  There is, therefore, some amount of consistency with other
approaches that criticize this bozonisation approach where excitons
are granted as individual particles (bosons) due to the impossibility
of ignoring particle exchanges. At least they dominate in the
approximation and can be reconciled to much extent with
bosonization~\cite{glazov09a}. Other microscopic terms of the
full-semiconductor polariton Hamiltonian are less problematic, in
particular phonon relaxation is well described with a Hamiltonian for
the coupled exciton-photon-phonon system modeled as coupling of the
deformation-potential of an electron-hole pair, to the lattice. The
calculations have been worked ou in detail by Pau \emph{et
  al.}~\cite{pau95b} and others~\cite{ciuti03a}.  Exciton interactions
become polariton interactions by multiplying them with the square of
the Hopfield excitonic coefficient modulus-squared~$|X|^4$. The
blueshift of a condensate with density~$n\equiv N/S$ is given by
$MN=gn$ with~$g$ the numerator of
Eq.~(\ref{eq:MonJan5062743PMCET2026}). For opposite
spins~\cite{vladimirova10a}, the biexciton should be included, in
which case another, and stronger, scattering channel becomes possible,
the Fesbach resonance~\cite{wouters07d,takemura14b}, which exhibits a
characteristic dispersive shape of resonant scattering with a rich
spin-dependence.  One of the constant preoccupations of polariton
physics is to further boost interactions.  Various routes can be
followed to do so, for instance involving high-orbital Rydberg
excitons, whose polaritons have been reported to also
condense~\cite{bao19a} while polariton-polariton interactions indeed
get substantially magnified due to the spatial extent of such
particles, that see each other over large distances~\cite{gu21a}.
Another strategy is to rely on indirect excitons, say in different
quantum wells or sheets of stacked 2D materials, and whose electron
and hole's spatial separation endows with a large dipolar moment.
They do not couple to light but can couple to polaritons, forming
so-called dipolaritons~\cite{cristofolini12a}, which can increase
their interaction strength by an order of
magnitude~\cite{datta22a}. There are many other approaches, for
instance, Kyriienko \emph{et al.}~\cite{kyriienko20a} suggest trions,
Tan \emph{et al.}~\cite{tan20a} polarons, while Bastarrachea \emph{et
  al.}~\cite{bastarracheamagnani21a} go directly for the full electron
gas.  The progress of material science, in particular with the
extension of graphene to a wide family of 2D materials, opened a new
chapter of polaritonics.  In contrast to traditional semiconductors,
2D materials are genuinely 2D, with no out-of-plane charges screening
of the Coulomb interaction, and can thus achieve binding energies in
excess of 500meV. Since this also makes for smaller bound states with
a reduced Bohr radius, their coupling with light also increases
significantly, resulting in much larger Rabi splitting. The recent
field of twistronics---exploiting band properties of 2D materials
stacked at an angle---brings in Moir\'e
nonlinearities~\cite{zhang21a}.

If we reduce the problem to a single-mode, say~$\mathbf{k}=0$ and
getting rid of this index, keeping particle-particle interactions and
driving of both modes, turns Eq~(\ref{eq:ThuJan1012203PMCET2026})
into:
\begin{equation}
  \label{eq:SunJan4104041PMCET2026}
  \begin{aligned}
    H&=\hbar\omega_a\ud{a}a+\hbar\omega_b\ud{b}b+\hbar g(\ud{a}b+a\ud{b})+{}    \\
    {}+&{U_a\over2}\ud{a}\ud{a}aa
    +{U_b\over2}\ud{b}\ud{b}bb
    +\Omega_a e^{i\omega_a t}a+\Omega_b e^{i\omega_b t}b+\mathrm{h.c.}
  \end{aligned}
\end{equation}
For generality, we consider here interactions of both the photon~$a$
and exciton~$b$ modes, although only excitons are assumed to interact,
in which case one simply sets~$U_a=0$, but photons might also
experience Kerr-type of nonlinearities, such as in photonic
molecules~\cite{liew10a}, in which case both coupled modes feature
interactions. We also consider driving both modes, although the usual
assumption is also that only one mode---the cavity---is driven, i.e.,
$\Omega_b=0$, but we have previously commented that experimentally,
both modes, exciton and photon, get actually driven by the laser, so
this generalization is more relevant. This simple Hamiltonian is found
in a broad variety of polaritonic problems, starting with squeezing
and polariton blockade.

\subsection{Squeezing \& Polariton blockade}

Bringing together quantum states of the condensate and interactions
takes us to the physics of squeezing and polariton
blockade.  
Linearity looks less exotic than interactions, although linear effects
may still account for the bulk of the phenomenology in interacting
systems, for instance through interferences, which are both subtle and
rich.  This has led to some disputes even in regimes based on
interactions~\cite{cilibrizzi14a,amo15a}. Regarding quantum
polaritonics, the approach has been broadly to rely on strong-enough
interactions to induce quantum effects. Two main ideas emerged from
this approach as far as condensates are concerned.  First, if a
condensate can be considered in its simplest approximation as a
coherent state, then ``squeezing'' can be considered as the simplest
approximation for a weakly-interacting condensate beyond the
mean-field description.  This arises by introducing quantum
fluctuations around the condensate. The quadratic Hamiltonian can be
diagonalized---the procedure being called in this case the Bogoliubov
transformation---which mixes particle creation and annihilation
operators for opposite-momentum pairs. This introduces to leading
order two-mode squeezing, with paired quasiparticles ejected from the
condensate.  Degenerate single-mode squeezing is less prominent as it
arises from higher-order fluctuations beyond the Bogoliubov
approximation. Because it is, however, simpler---restricting itself to
a single-mode as opposed to spreading over a continuum of pairs---we
focus here on this particular case. The quantum state is then a direct
counterpart of the coherent state~(\ref{eq:MonSep15092709PMCEST2025})
and is similarly obtained from the squeezing operator
${S}(\xi) \equiv \exp\left( ({\xi^* {a}^2 - \xi {a}^{\dagger 2}})/{2}
\right)$, where~$\xi\equiv re^{i\theta}$ is the squeezing parameter,
with $r$ the strength of squeezing (ellipticity of the deformed
coherent circle) and~$\theta$ its angle of squeezing. Applied to the
vacuum, this gives a squeezed vacuum state
$ |\xi\rangle \equiv {S}(\xi) |0\rangle = \exp\left(({\xi^* \hat{a}^2
    - \xi \hat{a}^{\dagger 2}})/{2} \right) |0\rangle$ whose Fock
state representation in the particle-number representation reads:
\begin{equation}
  \label{eq:MonJan5020231PMCET2026}
|\xi\rangle = \frac{1}{\sqrt{\cosh r}} \sum_{n=0}^{\infty} \frac{\sqrt{(2n)!}}{2^n n!} (-\tanh r)^n e^{i n \theta} |2n\rangle\,.
\end{equation}
This is a coherent superposition of pairs of particles and indeed the
terminology was initially that of a ``two-photon coherent
state''~\cite{yuen76a}. One can also squeeze the coherent state, i.e.,
squeeze a condensate, to reach a squeezed coherent state
$\ket{\alpha,\xi}$, which exhibits quadrature squeezing (reduced noise
in one quadrature below the coherent-state/vacuum level) around their
coherent amplitude which can be of high intensity (bright squeezed
state). This describes a condensate with degenerate squeezed
fluctuations. The state can be expressed as
$\ket{\alpha,\xi}=\sum_nc_n\ket{n}$ in terms of the Hermite
polynomials~$H_n$ in the coefficients
$c_n \equiv \big( \frac{\nu}{2\mu} \big)^{n/2} H_n
\big({\beta}/{\sqrt{2 \nu \mu}} \big) \exp(
-{|\beta|^2}/{2})/{\sqrt{n! \, \mu}}$ where $\mu \equiv \cosh r$,
$\nu \equiv e^{i\theta} \sinh r$ and
$\beta \equiv \mu \alpha + \nu \alpha^*$. This looks qualitatively
like a Poisson distribution but with sub-Poissonian statistics for
amplitude squeezing or general quadrature reduction.  Squeezing was
first considered for polaritons as early as 1993~\cite{hradil93a} and
subsequently given much theoretical consideration~\cite{tassone00a},
before the experimental report of first 4\%~\cite{karr04a} then
20\%~\cite{boulier14a} squeezing by the group of Elizabeth Giacobino,
in both cases through degenerate four-wave mixing by driving
resonantly the cavity at normal incidence, in which case the
single-mode Eq.~(\ref{eq:SunJan4104041PMCET2026}) holds. This can, in
a first approximation, be assumed for one lower-polariton operator
only as $H=\hbar\omega_\mathrm{LP}\ud{p}p+U\ud{p}\ud{p}pp$, making
obvious the connection between this degenerate parametric
configuration and single-mode squeezing~$S(\xi)$.  In this
configuration, squeezing is generated near the turning point of the
bistability curve.  The upgrade by Boulier \emph{et
  al.}~\cite{boulier14a} was thanks to the confinement of the
polariton modes in pillar microcavities, which shields them from the
dominant continuum of modes available for two-mode squeezing.
Although such resonant, degenerate parametric amplification is a quite
specific case, its demonstration of polariton squeezing has left a
lasting impression for going beyond the mean field of the condensate.

The second idea of nonlinearity-induced quantum effects of polaritons
is the so-called ``polariton-blockade'', which became the emblem of
quantum polaritonics, as going even beyond Gaussian (squeezing)
physics. This extends the concept of Coulomb blockade---whereby the
addition of a single electron to a small conductive island gets
suppressed by the large electrostatic charging energy required, thus
blocking conductance---to weakly-interacting (at the single-particle
level) polariton condensates.  Namely, when a coherent drive is in
resonance with the lower polariton branch---corresponding to the
one-particle energy---it gets out of resonance when the branch gets
populated, since self-interactions blueshift the state, which results
in a drop of the excitation now mismatched in energy.  This sieving of
the multiphoton components of the drive results qualitatively in
antibunching~\cite{gerace19a}. To be efficient with few particles,
interactions should be very strong, while we have observed that in
this regime, they remain fairly weak. Strong polariton blockade thus
remains a future target from progress in material science.  Its
theoretical proposal was first made by Cristiano Ciuti (corresponding
author for Verger and Carusotto)~\cite{verger06a} to study the
``quantum nonlinear dynamics'' of confined polaritons, which they
called ``photonic dot'' at the time. Using
Eq.~(\ref{eq:SunJan4104041PMCET2026}) with~$U_a=0$ (non-interacting
photons) and~$\Omega_b=0$ (driving the cavity), they swiped the
resonant excitation and studied the two-photon
correlator~$g^{(2)}(\tau)$, which, in the steady, reads
\begin{equation}
  \label{eq:WedJan7062759PMCET2026}
  g^{(2)}(\tau) = \lim_{t\to\infty}\frac{\langle \hat{a}^\dagger(t) \hat{a}^\dagger(t+\tau) \hat{a}(t+\tau) \hat{a}(t) \rangle}{\langle \hat{a}^\dagger(t) \hat{a}(t) \rangle^2}\,.
\end{equation}
The $\tau=0$ case corresponds to coincidences and typically captures
the strongest correlations, and is hence the default
observable. Antibunching---that is, $g^{(2)}(0)<1$---is a hallmark of
non-classicality, since it violates Cauchy-Schwarz inequalities
imposed on classical fields. Polariton blockade results in an
alternating bunching and antibunching around the lower polariton, as
shown in Fig.~\ref{fig:SunJan4111733PMCET2026}(a). This prediction has
long been regarded as the first milestone for quantum polaritonics. It
has been observed but with very small
antibunching~\cite{delteil19a,munozmatutano19a}. Would the seminal
theory~\cite{verger06a} have extended its ``pump detuning'' to the
full frequency window shown in
Fig.~\ref{fig:SunJan4111733PMCET2026}(b), however, it would have
reported a much stronger effect, later discovered by Tim Liew and
Vincenzo Savona in the different implementation of a photonic
molecule~\cite{liew10a}, where, instead of excitons and photons, one
has two interacting ($U_a\neq0$ and~$U_b\neq0$) cavity modes coupled
by tunneling.  The physics is the same and is part of the broader
picture that is revealed in Fig.~\ref{fig:SunJan4111733PMCET2026}(c):
a full two-photon landscape of the weakly interacting polariton
condensate. The case chosen is of very-strong polariton
interactions---to magnify the qualitative features---but those persist
even as the interactions become very
small. 
This landscape of correlations reveals, through multiphoton
observables, unsuspected mechanisms in the coupling of
coherently-driven oscillators in the presence of Kerr-type nonlinearities.
For each exciton-photon detuning on the~$x$ axis, the system's cavity
mode is resonantly driven by a laser at the frequency reported on
the~$y$ axis, and for all those cases, the cavity two-photon
statistics~$g^{(2)}(\omega_a,\omega_\mathrm{L})$, is color-coded,
namely, in red for bunching, blue for antibunching and zero for
uncorrelated light. In such a configuration, at the one-photon level,
one typically only observes the polariton branches, that resurface in
$g^{(2)}$ as the CA lines, which exhibit the familiar
anticrossing. This occurs when the laser probes the branches at
resonance.  In two-photon observables, however, much more polaritonic
structure is revealed. Importantly, this is regardless of the
intensity of the collected light, and one can observe two-photon
resonances in unsuspected regions where there is no manifestation of
signal at the one-photon level (i.e., no spectral peak or resonance).
The fact that two-photon correlations can be measured regardless of
how many photons are available to do so, is a general feature of
two-photon physics~\cite{zubizarretacasalengua24a} (this applies to
higher orders too). This differentiates a classical signal, which is
obtained from one-photon observables and is quantified by the
intensity, from a quantum signal, which is obtained from multi-photon
(starting with two-photon) observables and exist even in absence of
(much) classical signal.  The landscape of
Fig.~\ref{fig:SunJan4111733PMCET2026}(c) is interpreted in the CUBA
picture of Conventional/Unconventional
Bunching/Antibunching~\cite{zubizarretacasalengua20a} which explains
all the resonances resolved in the figure.  Conventional features are
those associated with dressed states (polaritons), while
unconventional ones are associated with interferences.  The LP and UP
branches, where the Conventional Antibunching (CA) of Verger \emph{et
  al.}~\cite{verger06a} is realized, feature a weak antibunching, but
with a strong classical signal (intensity). For this reason, it has
been the most pursued feature. The cavity population vanishes when it
is driven at the exciton energy, leading to superbunching, which has
been described as ``photon tunelling'', but it actually arises from an
interference that leaves only the quantum fluctuations of the Kerr
oscillator, which fits under the category of Unconventional Bunching
(UB). This is the bunching counterpart of the Liew \& Savona
Unconventional Antibunching (UA).  Both features---antibunching and
bunching---are indeed different manifestations from the same
interference of the coherent
state~(\ref{eq:MonSep15092709PMCEST2025})---induced by the laser
driving the system---with the squeezed vacuum
state~(\ref{eq:MonJan5020231PMCET2026}), induced by the weak Kerr-like
polariton nonlinearity. Although squeezing is small, a two-photon
observable is of this order anyway, and can lead to a perfect
suppression of the two-photon---or higher---component.  An important
characteristic of the unconventional blockade is that it can only
destructively interfere one component of the wavefunction at a time,
and is thus not a good mechanism for a single-photon source, which
should suppress all $N>1$-photon components. The other red lines in
Fig.~\ref{fig:SunJan4111733PMCET2026}(c) correspond to hitting the
two-polariton resonances and thus qualify better for ``tunelling'',
being this time a two-polariton resonance with the laser
out-of-resonance with the single-polariton branches, but resonant at
the two-photon level. This other conventional feature is, therefore,
the direct counterpart of polariton blockade, trading antibunching for
bunching and occurring elsewhere in the parameter space. We need not
discuss further here this two-photon picture of the weakly-interacting
condensate and we refer to its detailed CUBA analysis by Zubizarreta
Casalenga \emph{et al.}~\cite{zubizarretacasalengua20a,
  zubizarretacasalengua20b}. The point is that there is a consistent
and unified picture of the multiphoton (here, the two-photon) response
of a polariton BEC, whose famed polariton blockades---both
conventional and unconventional versions---are just particular cases
of a broader picture, that remains to be experimentally demonstrated,
as well as exploited. For instance, the crossing of the UA and CA
lines for the upper polaritons, promises record bright two-photon
antibunching as combining the best of both worlds: strong signal from
the emission from a branch and strong antibunching from the
interference.

\subsection{Quantum coherence}

A comparatively more recent development of the conceptual
interpretation of polariton condensation comes in the wake of the
quantum resource theory of quantum coherence~\cite{chitambar19a}.
Resource theory broadly emerged to quantify quantum advantage when
practical applications failed to become compelling on their own. It
started with a concern of quantifying/characterizing entanglement but
then got upgraded to other quantum resources, in particular
``coherence'', but in the sense of quantum superposition, or
off-diagonal terms in the density matrix.  An interesting aspect of
resource theories---which makes them increasingly important---is that
although they describe a broad variety of separate quantum properties,
they follow the same structure and can even relate some resources to
the others, e.g., connect quantum coherence to entanglement. They rely
on ``free states'' (which do not feature the resources, i.e., diagonal
states for coherence or product states for entanglement) and ``free
operations'' which are easy to implement and, in particular, cannot
create the resource on a state which does not possess it. For
coherence, the theory was initially developed for finite-dimensional
Hilbert spaces~\cite{baumgratz14a}, in particular, for qubits. In the
original sense, a superposition like (\ref{eq:Sun24Nov180218CET2024})
is coherent, while the reference states~$\ket{1,0}$ and~$\ket{0,1}$
are the free (easy to generate) states. The basic idea is that it is
easy to prepare an atom in its ground or excited state, which is
classical, but not to have it in a superposition, which is
quantum. Resource theories make such notions precise.

The quantum coherence of continuous variables, such as quantum states
of the harmonic oscillator, brings us to the conundrum that a
pure-state superposition like Eq.~(\ref{eq:MonSep15092709PMCEST2025})
becomes a quantum resource, i.e., it has quantum coherence (as the
name unfortunately further suggests) while the Fock states~$\ket{n}$
are the free states, with no quantum advantage or features. For
photons, however, Fock states are non-classical and thus associated to
quantum properties, while coherent states are classical, and thus
devoid of quantum advantage.  Even if including the light-matter
aspects---as we already discussed---the polariton Fock states are the
difficult states to produce, with so far only~$\kket{1}$ having been
realized~\cite{cuevas18a,suarezforero20a}, while the ``coherent
state'' is the classical state which follows simply from impinging a
laser onto the microcavity. Similarly, Fock states optimize quantum
correlations while coherent states cancel them. While this discrepancy
of terms has been noted (e.g., as a ``semantic
subtlety''~\cite{streltsov17a}, as ``notions of coherences in quantum
resource theory and quantum optics'' being ``fundamentally
different''~\cite{wu21a} or having to restrict the terminology ``to
avoid confusion''~\cite{killoran16a}), in polaritons, this
understanding of the coherent state as a quantum resource has been
embraced. Namely, the group of Marc~A{\ss}mann studied the problem of
polariton condensation but through the angle that the coherent
state~(\ref{eq:MonSep15092709PMCEST2025}) is a quantum resource.  In
this approach, they ``\emph{speciﬁcally consider a superposition of
  particles to represent quantum phenomena in matter systems}'' and
``\emph{identify particles as classical in order to provide a
  characterization of the matter system}''. Said otherwise, they
transpose to quantum resources the same discussion of how much the
light and matter parts of polaritons are accountable for their
properties, as in BEC, thus opening a new front for controversy.  The
quantifier of quantum coherence is defined for the density
matrix~$\rho$ as the sum of its off-diagonal elements:
$\mathcal{C}\equiv\sum_{n\neq m}|\rho_{m,n}|^2$~\cite{luders21a}, and
is thus zero if the state is diagonal, as is a thermal state, but also
a Fock state despite being a pure state.  While this approach to
quantum coherence has also been pursued outside of the polariton
community~\cite{zhang16b}---where the laser and superfluidity are
taken as examples of quantum macroscopic coherence---this brings about
weird features. For instance, the coherent state is not the most
quantum coherent one: the pure state whose diagonal elements are given
by the thermal distribution, is. This looks like a pathology of
inverting everything. A less confounding approach by Tan
\emph{et al.}~\cite{tan17a} introduces the~$\alpha$-coherence, that
more expectingly reconcile quantum resources of superpositions with
nonclassicality. In the polariton case, furthermore, the off-diagonal
elements themselves are not directly measured, as currently outside
the scope of today's tomographic capabilities, but are inferred
instead. The transition is assumed to go from the thermal
state~(\ref{eq:SunJan11112228AMWET2026}) below threshold to a coherent
one~(\ref{eq:MonSep15092709PMCEST2025}) above threshold, transiting by
a mixture of both in between, assumed to be a displaced thermal state
$\rho_\mathrm{coth}\equiv
D(\alpha)\rho_\mathrm{th}\ud{D}(\alpha)$~\cite{laussy06a}, with matrix
elements given by, for $m\ge n$:
\begin{multline}
  \label{eq:SunJan11121257PMWET2026}
  \langle m|\rho|n\rangle =\frac{1}{\bar{n}_\mathrm{th}+1} \exp\!\left(-\frac{|\alpha|^2}{\bar{n}_\mathrm{th}+1}\right)\times\\ \sqrt{\frac{m!}{n!}}\,(\alpha^*)^{m-n} L_n^{m-n}\!\left(-\frac{|\alpha|^2}{\bar{n}_\mathrm{th}}\right)
\end{multline}
where~$L_n^k(x)$ is the generalized associated Laguerre polynomial of
order~$k$ and degree~$n$.  The diagonal elements display a smooth
transition from an exponentially decreasing (thermal) to a peaked
(Poissonian) distribution, in agreement with direct measurements by
photon-counting methods~\cite{klaas18a}.  In contrast, while
Eq.~(\ref{eq:SunJan11121257PMWET2026}) indeed defines nonzero diagonal
elements, inherited from~$\ket{\alpha}$, those are more difficult to
access experimentally and L\"uders \emph{et al.}~\cite{luders21a}
calculated the quantum coherence~$\mathcal{C}=\rho_\mathrm{coth}$
indirectly, from measurements of the incoherent~$n_\mathrm{th}$ vs
coherent~$|\alpha|^2$ fractions, from a \emph{phase-averaged} Husimi
distribution~$Q$ (a representation of the density matrix which they
could measure by homodyning). In this way, they infer quantum
coherence from $\approx 0.2$~\cite{luders21a} up to above
$0.6$~\cite{brune25a} in annular-ring trapped condensates with long
lifetimes. Besides the traditional problem of coherence buildup from
the point of view of particle-number statistics (diagonal elements),
to which we come back in the following Section, L\"uders \emph{et
  al.}'s approach is also focused on the buildup of a well-defined
phase of the condensate. This was also the motive behind the author
\emph{et al.}'s definition of the order parameter of condensation as
the degree of linear polarization of the emitted
light~\cite{laussy06a}, with the underlying idea that if both circular
polarization components of the polariton ground state grow coherent
states with a well-defined phase, then they lock into a given linear
polarization. While this has received experimental
support~\cite{baumberg08a,delvalle09c} with time-resolved ultrafast
measurements, a fundamental difficulty had been raised by M{\o}lmer
who suggests that an absolute phase $\operatorname{Arg}(\alpha)$
in~$\ket{\alpha}$ is a ``convenient fiction'', in the sense that it is
a useful mental picture to undertake calculations on the basis of a
pure coherent state for a single-mode, which gives the same result as
what more likely takes place: entanglement with surrounding modes to
distribute the phase of a single mode into cross-correlators between
several modes instead. In M{\o}lmer's own words: ``\emph{it does not
  matter whether coherences exist or not; observable phenomena in
  optics and quantum optics are unchanged, and in this way optical
  coherences may be regarded as a convenient fiction.}''  In a
followup work, he comments that not only coherence might not be
needed, it might actually be impossible to
create~\cite{molmer97b}. The entanglement of single operators with all
other modes to which they couple, is extremely robust to disturbances
of each isolated mode. While this is complicated to keep track of, the
same equations follow from assuming a single mode in isolation whose
mean field is nonzero (a coherent state). This understanding remains a
point of dispute. Others, such as Carmichael \emph{et
  al.}~\cite{noh08a}, contend that, on the contrary, a laser does
produce a genuine coherent state that factorizes from its target
state. The problem of the existence of such a phase more broadly
belongs to a fundamental problem of condensation as a phase
transition, namely, the problem of symmetry breaking. By choosing a
phase---if it does---from an
Hamiltonian~(\ref{eq:ThuJan1012203PMCET2026}) which does not favour
any, the system breaks a global $U(1)$ gauge symmetry (continuous
rotation). This brings forward the problem of the
phase--particle-number complementarity: a system with a fixed number
of particles (a Fock state) has no phase defined, while when it
condenses (into a coherent state) it acquires one, also lifting the
strict conservation of particle numbers. A seminal observation was
made by Javanainen that a fixed-number of particles~$\ket{N}$ can
still acquire spontaneously a phase~\cite{javanainen91a}, which can be
demonstrated by interfering two BECs, each with a fixed number of
particles~\cite{javanainen96a}.  Polaritons are ideally suited to
revisit such fundamental aspects of the field, with tuneable degrees
of interactions that can bring those questions from purely interfering
optical states to strongly-interacting many-body atomic
ones. Javanainen's quantum phase and its symmetry breaking, for
instance, dispensed from interactions altogether. This could not only
clarify whether the phase is, indeed, irrelevant in such problems, but
also if quantum coherence might also be a fiction, although not even a
convenient one as bringing no simplification when actual resources are
to be quantified. A final comment is that bosonic correlations from
classical states---such as thermal states which display bunching---can
be understood as an artifact of assuming sampling from a given
geometry (a donut for vortices) while each correlated measurement is
actually performed in a different geometry (a dipole whose phase is
acquired by symmetry breaking), where they are sampled as uncorrelated
particles~\cite{arXiv_salazar26a}. This shows that bunching from
classical states is not as mysterious as one could imagine, like from
Fock states, which do feature genuine superpositions of all possible
dipole orientations, simultaneously, which could be processed
quantum-coherently. Such fundamental questions are all within the
scope of the physics of polariton condensates, extending the problem
of their nucleation to all elements of the density matrix.

\section{Nucleation of the Polariton condensate}

\subsection{Boltzmann Equations}

Given the short polariton lifetime, polariton condensation is an
intrinsically dynamical problem, or at least one out-of-equilibrium,
when balanced by continuous pumping.  The \emph{formation} (or
``nucleation'') of the polariton BEC is thus a central problem of
polariton condensation. For atoms, in contrast, this is a niche topic
in addition to being an extremely complex one, that produced few
well-established results~\cite{stoof92a}. This is another arena where
polaritons can claim a privileged access to the physics of BEC and
where, at least, considerable activity has been and keeps being
expanded.  As a blueprint of---or objective for---their steady-state,
one has the ideal (non-interacting) polariton gas, in which case the
Hamiltonian is simply Eq.~(\ref{eq:ThuJan1013440PMCET2026}), which we
can even further simplify by assuming only lower polaritons. The
Bose--Einstein condensation theory of this system from standard
quantum thermodynamics provides the density matrix in equilibrium at
temperature~$T$ and with chemical potential~$\mu$ as
$\rho = \frac{1}{\mathcal{Z}} e^{-\beta (H - \mu N)}$ where
$\mathcal{Z}$ is the grand partition function with
$\beta\equiv 1/(k_\mathrm{B}T)$ the Boltzmann constant and
$N \equiv \sum_k p_k^\dagger p_k$ the total polariton-number
operator. Since the modes are decoupled, the density matrix is a
product state $\rho = \bigotimes_k \rho_k$ and one recovers the same
qualitative result as for massive particles (atoms or molecules) for
the particle distributions
$\langle n_k \rangle = \operatorname{Tr}(\rho_k n_k) =
\frac{1}{e^{\beta (E_\mathrm{LP}(k) - \mu)} - 1}$ at the only, but
notable, exception that one has the polariton
dispersion~(\ref{eq:ThuJan1013440PMCET2026}) rather than the quadratic
one of massive particles. Phase diagrams have been computed on this
basis~\cite{malpuech03b}.  The simplest dynamical picture to study how
well can polaritons approach the Bose equilibrium density matrix is
that of kinetic equations for the mean occupation of each available
point in phase space. The main theoretical apparatus to attack this
problem is given by the Boltzmann equations (or transport equations),
that are kinetic equations describing the dynamics of a gas or fluid
in its phase space, i.e., not the whereabouts of each and every
particle but the probability that a given configuration for any one of
them be realized.  For a uniform system, one can consider reciprocal
space only and ignore motion in real space, to focus on the
distribution of momentum. In their simplest formulation, such
equations have the intuitive character of rate equations for the
populations~$n_\mathbf{k}$ of each mode with wavevector~$\mathbf{k}$
which one can understand from the physics of coupled sinks, i.e., how
much containers fill up or empty under their combined inputs and
losses:
\begin{equation}
  \label{eq:SatAug30124841AMCEST2025}
  \partial_t n_\mathbf{k}=P_\mathbf{k}-\gamma_\mathbf{k} n_\mathbf{k}-n_\mathbf{k}\sum_\mathbf{l}W_{\mathbf{k}\to \mathbf{l}}+\sum_\mathbf{l}W_{\mathbf{l}\to \mathbf{k}}n_\mathbf{l}\,.
\end{equation}
This set of coupled differential equations can be written down
directly on physical grounds: the rate of change of the population for
an uncoupled mode~$\mathbf{k}$ increases first as the rate of
pumping~$P_\mathbf{k}$, as this is precisely how much additional
particles are being brought per unit time, and also decreases due to
the decay rate~$\gamma_\mathbf{k}$ that is, however, weighted by the
population~$n_\mathbf{k}$, as each particle has, independently from
the others, the possibility to radiatively decay. Coupling the
mode~$\mathbf{k}$ to all others~$\mathbf{l}$ redistribute the
particles between the states depending on their scattering rates,
first from the mode itself towards others, at a
rate~$W_{\mathbf{k}\to\mathbf{l}}$ resulting in a decrease of the
population, as well as, the inverse process whereby other modes act as
a pump term, in proportion to their respective available populations
and scattering rates. We would not discuss in such detail such
trivial equations---that are straightforwardly solved
analytically---if not for the fact that in the same spirit, we can
upgrade them to one of the most important set of equations for
polariton BEC, known under a variety of names, from the quantum
Boltzmann equation to the Uehling--Uhlenbeck
equation~\cite{uehling33a}. This is done by adding the basic, but
fundamental, Bose stimulation:
\begin{multline}
  \label{eq:SatAug30012756PMCEST2025}
  \partial_t n_\mathbf{k}=P_\mathbf{k}-\gamma_\mathbf{k} n_\mathbf{k}\\-n_\mathbf{k}\sum_\mathbf{l}W_{\mathbf{k}\to \mathbf{l}}(1+n_\mathbf{l})+(1+n_\mathbf{k})\sum_\mathbf{l}W_{\mathbf{l}\to \mathbf{k}}n_\mathbf{l}\,.
\end{multline}
One goes from the trivial Eq.~(\ref{eq:SatAug30124841AMCEST2025}) to
the analytically intractable Eq.~(\ref{eq:SatAug30012756PMCEST2025})
by increasing the scattering rates, say, from
$W_{\mathbf{l}\to\mathbf{k}}$ previously to now
$W_{\mathbf{l}\to\mathbf{k}}(1+n_\mathbf{k})$, a function of the
occupation of the final state. From a quantum perspective, the term
${}+1$ describes spontaneous scattering, whereby the particles arrive
by chance regardless of the final state, which is the classical
process from the previous equation, while $n_\mathbf{k}$ describes
Bose stimulation that magnifies the scattering towards populated
states: the bunching effect.  This is the type of equation proposed by
Fr\"ohlich~\cite{frohlich68b}, with constrained coefficients, and
which have been adapted or refined to capture a number of
specificities~\cite{yura94a,imamoglu96a, imamoglu96b}.  There are
various ways to derive such equations, in particular from the
microscopic Hamiltonian~(\ref{eq:ThuJan1012555PMCET2026}),
supplemented with a relaxation mechanism. The above type of rate
equations follows typically from phonon-mediated relaxation.  As such,
this describes the ideal gas, which can thermalize with its
surrounding but does not self-interact.  One can also include
polariton-polariton interactions Eq.~(\ref{eq:MonJan5061854PMCET2026})
which yield equally transparent rate equations of the
type:~\cite{tassone99a,doan08a}
\begin{multline}
  \label{eq:SatAug30025540PMCEST2025}
  \partial_tn_\mathbf{k}=\partial_tn_\mathbf{k}\Big|_\mathrm{Eq.\,(\ref{eq:SatAug30012756PMCEST2025})}-\sum_{\mathbf{l},\mathbf{k}',\mathbf{l}'}W_{\mathbf{k},\mathbf{k}',\mathbf{l},\mathbf{l}'}\big[n_{\mathbf{k}}n_{\mathbf{l}}(1+n_{\mathbf{k}'})(1+n_{\mathbf{l}'})\\
  +n_{\mathbf{k}'}n_{\mathbf{l}'}(1+n_{\mathbf{k}})(1+n_{\mathbf{l}})\big]\,.
\end{multline}
Other processes have been proposed as well, for instance substituting
phonons for free electrons~\cite{malpuech02a}.  Such microscopic
derivations provide first-principles expressions for the scattering
rates.  No phase information is retained as the relaxation proceeds
from independent collisions due to scattering derived from Fermi's
golden rule.  As such, the most fitting name for those equations
is---rather than ``quantum'' or ``Uehling--Uhlenbeck''---the other
popular qualification of ``semi-classical'' Boltzmann equations.  A
recurrent theme of such Boltzmann descriptions is that although they
only involve populations, they remain extremely complex, requiring one
to take into account a full two-dimensional phase-space for various
entities (polaritons, phonons, free carriers, etc.) that further vary
strongly depending on which part of the dispersion they are taken from
(photon-like and exciton-like polaritons), with also strongly
momentum-dependent scattering elements, ill-defined or little
understood self-interactions, the presence of various types of
disorder and dephasing, finite lifetime, pumping, etc. The list of
possible solid-state refinements is unending.  Nevertheless, despite
such forbidding complications, the phenomenology often appears to be
well captured by simple toy models. This is illustrated already from
one of the earliest such treatments, by Pau \emph{et
  al.}~\cite{pau95b}, who derive the rate equations from the
microscopic Hamiltonian, provide extensive details on the scattering
rates, solve the whole system numerically, to finally compare their
experimental results to an effective toy-model only featuring two
populations: ``dressed excitons'' (interestingly, like Weisbuch, the
authors do not want to call the~$k=0$ states polaritons as they do not
propagate) and ``propagating polaritons''. They find from both their
analytical results (of the toy model) and their experiments, a
threshold behaviour for population buildup with pumping which is
temperature dependent and thus departing from conventional lasing.
Such a dynamical description of the polariton population is needed
even in absence of BEC, to describe the luminescence of a system that
acquires all the properties of a gas.  BEC is ``merely'' the most
interesting and exotic aspect of this problem. If $n_\mathbf{l}\ll 1$,
then the much simpler Eq.~(\ref{eq:SatAug30124841AMCEST2025}) applies
and can still reproduce features of interest, such as bi-exponential
decay of time-resolved photoluminescence~\cite{bloch97a}. Many groups
have tackled the relaxation dynamics, and even though not all were
concerned about condensation, this aspect has often been lurking in
the background. For instance, one of the most important results of the
polariton Boltzmann equations is the identification by Tassone
\emph{et al.}~\cite{tassone97a} of a strong bottleneck effect in the
relaxation assisted through phonons alone, where they get ``stuck''
and accumulate in intermediate energy states on their way towards the
ground state. This is due to the dispersion turning from flat at high
momentum (with a high density of states) to parabolic near the ground
state (with a density of states dropping to a constant proportional to
the small polaritonic mass). This makes polaritons accumulate around
the inflection point of the dispersion where they change character
from exciton- to photon-like, as they cannot easily find
energy-momentum matching phonons to make this transition.  This
bottleneck effect was already known from bulk polaritons, and its
manifestation in cavity polaritons was initially received as an
intrinsic limitation of such systems. However, although a hurdle for
relaxation, this isolation of the ground state is actually an asset
for condensation, and we already encountered the decrease of the
density of states as responsible for the high critical
temperature. Despite, or due to, the bottleneck, Boltzmann equations
can show large occupations $n_\mathbf{k}\gg1$ at finite momenta~$k$ of
the dispersion, which is furthermore typically assumed isotropic
(all~$k\equiv|\mathbf{k}|$ states are populated the same) for
convenience and in agreement with fast elastic scatterings present
experimentally. This makes the process towards polariton condensation
take the peculiar form of rings in the far field (detection as a
function of angle) with shrinking diameters as pumping is
increased~\cite{savvidis02a}. More and more particles accumulate
closer to the ground state, although still in highly degenerate states
so there is no macroscopic occupation of a single mode. However, such
a dramatic buildup of a piling-up reservoir ready to collapse into the
ground state invites us to push further to overcome this bottleneck.
The most obvious theoretical solution---polaritons with longer
lifetimes---was not immediately practical in the laboratory, but
various other ways have been found, first of all, the presence of
polariton-polariton interactions to boost and complement phonon
relaxation. This was studied by Porras \emph{et al.} (following and
extending Tassone \& Yamamoto's similar approach~\cite{tassone99a})
who thus reproduced the rings in $k$-space~\cite{porras02a}, as well
as other experimental observations such as the negligible population
from the upper polariton branch, in agreement with
observations~\cite{lesidang98a} (theory was in fact initially
predicting dominant upper polariton emission~\cite{tassone97a}).
Malpuech \emph{et al.} suggested doping with a free electron gas for
an expeditive improvement~\cite{malpuech02a}. There are still other
ways, including with phonons alone, such as merely distorting the
polariton dispersion with detuning to weaken the phonon
bottleneck~\cite{soroko02a} and merely by pumping strongly
enough~\cite{doan02a}, which does not work in the bulk even if
extending the lifetime but can overcome the bottleneck in 2D, at least
theoretically. One way or another, polaritons have succeeded in
overcoming the dissipative battle and to condense into their ground
state. Although the exact nature of the object thus formed remains a
complex one opened to debate, it is unequivocally accepted that one
can pile up a macroscopic number of polaritons into their ground
state.







\subsection{Macroscopic ground state population}

The main feature of BEC remains the macroscopic occupation of the
ground state.  Since Yura and Hanamura's first theoretical prediction
of this possibility for polaritons, various groups have reported
macroscopic occupation of the ground state~\cite{porras02a}. The best
and more detailed efforts within that framework are from David Snoke
\emph{et al.}, who confronted quantitatively their model to
experimental results. Snoke is an expert on such problems, having
already provided direct and pioneering numerical fitting of
experimental time-resolved exciton dynamics in Cu$_2$O. In the case of
polaritons, he provided realistic models, including polariton-phonon,
polariton-polariton but also polariton-electron scattering channels
which was found to be necessary to account for the observations. This
was not injected on purpose---as suggested by
Malpuech~\cite{malpuech02a}---but was present from the stress in the
sample. Snoke \emph{et al.} also relaxed the previous approximations
of separating the populations between polaritons in their trap and
excitons in a flat band acting essentially like a reservoir. Instead,
they considered the gas as a whole, confirming, rather than assuming,
a decoupling of two populations. They even implemented a mechanism of
renormalization of the dispersion due to interactions, that is beyond
the scope of Boltzmann equations. They found a great agreement with
the experiment and gained insights such as the need of chopping the
laser and avoid over-heating the sample to allow a relaxation in
agreement with their Boltzmann dynamics. In such a case, they also
found large occupation of the ground state indeed, but without
thermalization of the high-energy states, confirming the
out-of-equilibrium character of polariton condensation.  This
confirmed a phenomenology alluded to by many but best systematized by
Kasprzak \emph{et al.}~\cite{kasprzak08b} of a thermodynamic vs a
kinetic regime of condensation, who, to be precise, further break the
kinetic regime in two, to arrive at this classification:
\begin{itemize}
\item \emph{thermodynamic condensation}: with thermalization both below and above threshold, and condensation ruled by thermodynamics, like atomic BEC.
\item \emph{kinetic condensation}: with no thermalization of the distribution but condensation in the ground state.
\item \emph{kinetically driven Bose condensation}: with a transition
  from a state with no temperature below threshold to a condensate
  with a temperature above threshold.
\end{itemize}
Kasprzak \emph{et al}'s nomenclature was also compared to experiments
and found to be ruled by two key parameters: detuning and the lattice
temperature. Thermodynamic condensation occurs at positive detuning
(photon-like polaritons in the ground state) while kinetic
condensation is at negative ones (exciton-like), which is in agreement
with \Imamoglu's early picture of the phenomenon.  There are various
additional insights which are worth noting: the kinetically-driven BEC
occurs in a narrow range of parameters, and kinetic regimes strongly
rely on polariton-polariton interactions to overcome the bottleneck
(called here the kinetic barrier). In contrast, the thermodynamic
regime can occur with phonons alone, and although the equilibrium
temperature is typically higher than the lattice, polaritons can also
equilibrate with the lattice if they have longer lifetimes, in which
case dissipation plays a minor role and the atomic scenario is
recovered. Another important difference is that the kinetic regime is
helped by higher lattice temperatures---speeding-up relaxation---while
the thermodynic one favours lower temperatures, in agreement with the
conventional picture of BEC. The authors also find a resolution to the
BEC vs lasing polariton dilemma, at least within the Boltzmann scope:
while the ``polariton BEC'' terminology is exclusive to the
thermodynamic regime, ``\emph{the condensate emits coherent photons
  and can be certainly called a polariton laser, if one is interested
  in fabricating a device.}'' Pragmatism wins. This picture received
its final touch from Snoke, with the report of completely equilibrated
polariton distributions in samples where they have long-enough
lifetimes~\cite{sun17a}, in agreement with theoretical
predictions. The group made cavities with $Q$ factors
of~$\approx 320,000$ resuling in polariton lifetimes of 270ps, two
orders of magnitude higher than in the early days.  This was achieved
thanks to better mirrors by increasing the number of layers in the
distributed Bragg reflectors of the cavity. Further trapping the
condensate in an optical trap---which favours condensation and also
provides a spatially-separated excitonic reservoir sitting on the
shoulders of the trap---they isolate a clean polariton condensate
which fits the Bose-Einstein distribution up to~$1.1$ times the
condensation threshold, beyond which the occurence of the condensate
breaks down their Boltzmann treatment.  They attribute such mismatch
to the expected $\delta$-function peak being broadened by finite-size
fluctuations, which spread over the rest of the
distribution. Remarkably, however, the tails of the distributions are
cut off as particles accumulate in the low-energy states. This overlap
indicates that the gas has reached saturation: any additional
particles occupy the ground state, while the population of excited
states stops increasing. This is precisely the Einstein scenario of
Bose condensation. They also reconstructed from their experimental
data the linear relationship between the density of polaritons~$n$ and
their fitted temperature~$T$, expected at threshold from relating the
distance~$r_s\approx1/\sqrt{n}$ between polaritons to the thermal de
Broglie wavelength~$\lambda_T\approx\hbar/\sqrt{m k_\mathrm{B}T}$,
showing convincingly that the condensate builds up as polaritons start
to overlap as waves. With such demonstrations of a macroscopic
population buildup in the ground state and redistribution of the rest
of particles according to the Bose-Einstein distribution, one has been
as far as Boltzmann equations can go. The condensate itself, however,
is expectedly much more complicated and rich than its mere population,
and the polariton version differs even more from its atomic paradigm
than it does with its mechanisms of relaxation. One needs, however, to
either extend or replace Boltzmann equations.




\subsection{Quantum theory of polariton condensation}
\label{sec:SatAug30024940PMCEST2025}

The most severe limitation of the semi-classical Boltzmann equations
is that they remain intrinsically classical: the ``semi'' upgrade
consists of a self-consistent renormalization of the scattering rates,
to take into account Bose stimulation of the populations. While this
still results in interesting departures from the fully classical rate
equations~(\ref{eq:SatAug30124841AMCEST2025}), this leaves aside
quantum aspects such as the wavefunction, quantum correlations or
nonclassical dynamics like superfluidity. There is a deeply ingrained
belief that particle interactions are crucial. We now delve in some
amount of details into what, precisely, interactions may contribute to
the problem.  As for basically all questions that can be put to
polaritons, one can try to adapt the efforts from older and bigger
research communities, such as that of cold atoms, to the specificies
of polaritons.  Sarchi and Savona undertook a substantial effort in
that direction~\cite{sarchi07a} by bringing to polaritons the
Hartree-Fock-Bogoliubov (HFB) formalism, that consists in decomposing
the field operators~$\hat p_{\mathbf{k}}$ into a condensate
component~$P_{\mathbf{k}} \hat a$ (here~$\hat a$ is the ground state
operator and $P_{\mathbf{k}}$ a scalar amplitude) and the fluctuations
around that~$\tilde p_{\mathbf{k}}$. Since we are mixing quantum
operators and~$c$ numbers, we are following these authors' notation
and use hats for operators. The decomposition thus reads
$\hat p_{\mathbf{k}}\equiv P_{\mathbf{k}}\hat a+\tilde
p_{\mathbf{k}}$, for all~$k$.  This accounts for both the condensate,
which provides an order parameter for the transition, and the
non-condensate, that describes the excitations of the interacting
system. This treatment is further made in the so-called Popov
approximation, which is ad hoc for BEC by neglecting higher-order
terms to ensure a gapless excitation spectrum and thus satisfy the
Goldstone theorem.  Kinetic equations are derived in this way and
correct the semi-classical Boltzmann
equations~(\ref{eq:SatAug30012756PMCEST2025})
or~(\ref{eq:SatAug30025540PMCEST2025}) as follows:
\begin{subequations} 
  \label{eq:TueSep2081743PMCEST2025}²
  \begin{align}
    \partial_t n_c&=\partial_tn_c\Big|_{\mathrm{Eq.\,(\ref{eq:SatAug30025540PMCEST2025})}}+{2\over\hbar}\sum_{\mathbf{k}}v_{-\mathbf{k},\mathbf{k}}^{(\mathbf{q})}\operatorname{Im}(\tilde m_\mathbf{k})\,,\\
    \partial_t\tilde n_{\mathbf{k}}&=\partial_t\tilde n_{\mathbf{k}}\Big|_{\mathrm{Eq.\,(\ref{eq:SatAug30025540PMCEST2025})}}-{2\over\hbar}v_{-\mathbf{k},\mathbf{k}}^{(\mathbf{q})}\operatorname{Im}(\tilde m_\mathbf{k})\,,
  \end{align}
\end{subequations}
where~$n_c=\langle\ud{a}a\rangle$ is the ground-state population where
we expect the BEC and~$\tilde n_\mathbf{k}$ are the quantum
fluctuations of the excited states. Assuming that the condensate takes
the simple form of a plane wave in the ground state, i.e.,
$P_{\mathbf{k}}=e^{i\phi}\delta_{\mathbf{k},0}$,
then~$\hat p_{\mathbf{k}}=\tilde p_{\mathbf{k}}$ for the excited
states, and what would be otherwise the population of quantum
excitations of the collective modes from the interacting system comes
back to the population of the polaritons as in the previous
sections. If that wouldn't be the case, then
$n_\mathbf{k}=\tilde n_{\mathbf{k}}+|P_{\mathbf{k}}|^2 n_c$ and the
condensate spills over the excited states, but in this approximation,
the main addition reduces to the quantum correlator
\begin{equation}
  \label{eq:TueSep2085326PMCEST2025}
  \tilde m_{\mathbf{k}}\equiv \langle\ud{a}\ud{a}\tilde p_{\mathbf{k}}p_{-\mathbf{k}}\rangle
\end{equation}
which is the two-body correlations corresponding to two-particle
scatterings between the condensate and the excited states, with
opposite momenta.  Those are non energy-conserving processes that
cannot be treated within the Boltzmann framework. To close the set of
equations~(\ref{eq:TueSep2081743PMCEST2025}), one needs also the
equation for $m_{\mathbf{k}}$, but this opens an infinite hierarchy,
that involves, for instance~$\langle\ud{a}\ud{a}aa\rangle$ the ground
state coherence and squeezing operator on the one hand,
and~$\langle\ud{\tilde p}_\mathbf{q}{\tilde p}_\mathbf{-q} {\tilde
  p}_\mathbf{k}{\tilde p}_\mathbf{-k}\rangle$ the two-body
correlations between excitations on the other hand. This is where the
Popov truncation scheme comes to finally provide a complex nonlinear
set of equations but that involves only~$n_c$, the $n_\mathbf{k}$ (and
the exciton reservoirs as a Boltzmann decoupling of the coherent and
incoherent dynamics, for convenience), as well
as~$\langle\tilde m_\mathbf{k}\rangle$. The scattering rates must be
dynamically computed too, and they furthermore now depend not on the
polariton dispersion but on its renormalized Popov version. This is,
however, only a numerical matter and we focus instead on the results,
as reported by Sarchi and Savona. The main contribution is that
of~$\operatorname{Im}(\tilde m_1)$, the scattering with the closest
set of excited states, which is strongly negative and thus depletes
the condensate, which the authors say is more pronounced than for
atomic gases. They also see this in the off-diagonal long-range order,
which their formalism gives access to, and find it constant at large
distances, indicating a finite condensed fraction, but smaller than
for atomic gases. The distribution of the excitations in energy
expectedly does not follow the Bose--Einstein distribution, but also
deviates strongly from the equilibrium (atomic) Popov one. This could
be compatible with the beyond-threshold Sun \emph{et al.}'s departures
from Bose-Einstein distributions. The dispersion in momentum becomes
linear at small exchanged momentum, but the HFB Popov approximation is
precisely tailored to achieve that. The theory has little predictive
power, since several of its predictions can also be explained by less
sophisticated models. For instance, it predicts a decrease of the
condensate fraction with increasing condensate area, since it depends
on discretization of the states around the condensate, until, for
infinite-size systems, the condensate is completely lost, which the
authors see as a manifestation of the Hohenberg-Mermin-Wagner
theorem. This, however, is the same prediction made with Boltzmann
equations, essentially for the same reasons~\cite{doan08a}.  We have
chosen Savona's input to illustrate this approach to the problem,
given his long-time expertise and dedication to polaritons, but such
microscopic models have been similarly studied by other groups and one
could say that such treatments suffer from theory that is extremely
complex and sophisticated, requiring very heavy numerical simulations,
but only producing quantitative corrections which make it difficult to
assess what it truly captures that goes beyond semi-classical or
phenomenological models. A more popular and more broadly studied
approach is that of the Marburg school of Kira and Koch, and their
cluster expansion for handling correlations in interacting particle
systems. This involves decomposing quantum averages of operator
products (observables) into correlated ``clusters'' (singlets,
doublets, triplets, etc.), allowing for a systematic truncation of the
infinite hierarchy of equations of motion that arise from the
Heisenberg equations. The higher-order correlators get factorized into
lower-order terms plus fluctuations, enabling closed-form
approximations up to some order in the truncation. This allows to
capture effects like screening, carrier thermalization, bound states,
etc., as well as the breakdown of the polariton quasi-particle
pictures to melt into an excitonic plasma bathing in light.  Such
approaches have been instrumental in the early days, although in a
critical way to oppose the earliest interpretations of polariton
condensation~\cite{kira97a}, even getting to a point of challenging
the one-exciton picture \cite{kira98a} (let alone exciton-exciton
scattering), with the claim that the excitonic resonance in
luminescence stems from correlation effects from a plasma with no
excitons.  More favorable towards polariton BEC but still relying on
sophisticated many-body, diagrammatic and other advanced theoretical
picture, the Cambridge school from Littlewood's group contributed a
plethora of works~\cite{eastham01a, szymanska03a, keeling04a,
  marchetti04a, keeling05a, marchetti06a, szymanska06a}. Still, such
apices of theoretical edifices failed to provide clear-cut, compelling
and highly predictive models.


\subsection{Quantum Boltzmann Master Equations}



Even the Off-Diagonal Long Range Order is intrinsically classical. If
one wants to make a genuinely quantum characterization of the
polariton fluid, one needs quantum observables. This requires quantum
kinetic equations, although the other way around is not necessary, as
demonstrated for instance by Sarchi \& Savona's
treatment~\cite{sarchi07a} which is quantum but only computes
populations. They could, of course, also compute quantum observables,
but their treatment is so complex that they stopped at the quantum
quantities not easily or directly observables, but that are needed to
obtain classical observables as modified by the quantum dynamics. In
this way, they have all the inconveniences of a quantum treatment
(complexity of treatment) but none of its advantages (quantum
phenomenology). Here, we take the other end of this problem by
describing the simplest extension of the Boltzmann equations that
allows us to measure quantum observables. This does not consist of
coupling the dynamics to new correlators, such as
Eq.~(\ref{eq:TueSep2085326PMCEST2025}), and can indeed be fruitfully
reduced to the simplest formulation of condensation which is that of
phonon-assisted relaxation only, which is known to be able to drive
macroscopic population buildup on its own, as discussed
previously. The paradigm change is to compute the dynamics
of~$n_\mathbf{k}$ as random variables rather than, as was the case
before, as mean values~$\langle n_\mathbf{k}\rangle$. 
Furthermore, partly due to the fictitious nature of the phase as well
as inaccessibility of the purity of the state, it is also enough to
stick to the diagonal elements of the many-body wavefunction
$p(n_\mathbf{0},n_{\mathbf{k}_1},\cdots,n_{\mathbf{k}_N},\cdots)$
where those variables take all possible integer values. The basic idea
of a quantum Boltzmann master equation is that one retain correlations
between particle numbers, so
$\langle n_{\mathbf{k}_i}n_{\mathbf{k}_j}\rangle$ does not have to be
equal to
$\langle n_{\mathbf{k}_i}\rangle\langle n_{\mathbf{k}_j}\rangle$ as in
the Boltzmann case. This already points to a quantum aspect since in
this way we are considering correlations between states. In
particular, applied to the ground state itself, this brings us to
contrast $\langle n_0^2\rangle$ to $\langle n_0\rangle^2$, i.e., to
the variance of the population. In quantum optics, following Glauber,
it is traditional to measure this through  $g^{(2)}$, defined in
Eq.~(\ref{eq:WedJan7062759PMCET2026}).  In this way, we can provide a
dynamical derivation, i.e., in terms of quantum rate equations, of the
central quantum statistical object, that is the density matrix~$\rho$
(at least of its diagonal elements). The inclusion of the bath can be
done with dissipative coupling, i.e., with a master equation for the
density matrix in the Lindblad
form~$\partial_t\rho=-i[\rho,H]+\mathcal{L}\rho$ where the
superoperator~$\mathcal{L}\equiv\mathcal{L}_\mathrm{decay}+\mathcal{L}_\mathrm{pumping}+\mathcal{L}_\mathrm{phonons}$
brings in familiar terms in dissipative quantum optics, describing
decay~$\mathcal{L}_\mathrm{decay}\rho\equiv\sum_{\mathbf{k}}\frac{\gamma_\mathbf{k}}{2}(2p_\mathbf{k}\rho\ud{p}_{\mathbf{k}}-\rho\ud{p_\mathbf{k}}p_\mathbf{k}-\ud{p_\mathbf{k}}p_\mathbf{k}\rho)$
where~$\gamma_\mathbf{k}$ is the decay rate of the polariton with
corresponding momentum, e.g., neglecting the exciton lifetime,
$\gamma_\mathbf{k}=\gamma_c|C_\mathbf{k}|^2$ is simply the photon
decay rate weighted by the photon fraction
$|C_\mathbf{k}|^2$. Incoherent pumping does the reverse process of
randomly adding a particle of momentum~$\mathbf{k}$ at
rate~$P_\mathbf{k}$ with
$\mathcal{L}_\mathrm{pumping}\rho\equiv\sum_{\mathbf{k}}\frac{P_\mathbf{k}}{2}(2\ud{p}_\mathbf{k}\rho{p}_{\mathbf{k}}-\rho{p_\mathbf{k}}\ud{p}_\mathbf{k}-{p_\mathbf{k}}\ud{p}_\mathbf{k}\rho)$
with~$P_\mathbf{k}$ depending on the pumping configuration, e.g.,
normally distributed at high~$\mathbf{k}$ in the excitonic
region. Lesser known is dissipative coupling, transferring
polaritons from momentum~$\mathbf{k}$ to~$\mathbf{q}$,
\begin{equation}
  \label{eq:ThuJan8122822AMCET2026}
  \mathcal{L}_{\mathbf{k}\to\mathbf{q}}\rho\equiv{w_{\mathbf{k}\to\mathbf{q}}\over2}\big(
  2p_{\mathbf{k}} p_{\mathbf{q}}^{\dagger} \rho p_{\mathbf{q}} p_{\mathbf{k}}^{\dagger} -  \rho p_{\mathbf{q}}^{\dagger} p_{\mathbf{k}} p_{\mathbf{q}}^{\dagger} - p_{\mathbf{q}} p_{\mathbf{k}}^{\dagger} p_{\mathbf{k}} p_{\mathbf{q}}^{\dagger} \rho
  \big)\,,
\end{equation}
at the rate~$\omega_{\mathbf{k}\to\mathbf{q}}$, which can be, for
instance, phonon assisted. This is in the Lindblad (dissipative) form
because the phonon bath itself has been traced over, so correlations
with the phonon bath is not retained, \emph{but} correlations between
polaritons in the various modes are.  The population in
mode~$\mathbf{q}$ increases as a result of decreasing it in
mode~$\mathbf{k}$ but not only on average---as is the case with
Boltzmann equations---but in a correlated way. Such correlations are
sufficient to grow coherence of the condensate beyond a mere large
population, but in the quantum sense of Glauber of transitioning from
a thermal to a coherent state. The full relaxation in the polariton
gas as a whole follows from
$\mathcal{L}_\mathrm{phonons}\rho\equiv\sum_{\mathbf{k}\neq\mathbf{q}}\mathcal{L}_{\mathbf{k}\to\mathbf{q}}$,
though to understand the underlying mechanism, it is worth considering
the simplest case that exhibits the basic phenomenology, which is as
simple as two dissipatively coupled oscillators~$a$
and~$b$~\cite{laussy04a}, in which case
Eq.~(\ref{eq:ThuJan8122822AMCET2026}) reduces to
$\mathcal{L}={w_{a\to b}\over2}\mathcal{L}_{a{b}^{\dagger}}+ {w_{b\to
    a}\over2}\mathcal{L}_{{a}^{\dagger}b}$ and, even ignoring pumping
and decay for now, one gets to the following quantum Boltzmann (rate)
equations for the joint
probability~$p(n,m)\equiv\bra{n,m}\rho\ket{n,m}$ to have~$n$ particles
in mode~$a$ (the ground state) and~$m$ in mode~$b$ (the excited
state):
\begin{equation}
  \label{eq:WedJan7063208PMCET2026}
  \begin{aligned}
\dot p(n, m) &= (n + 1)m \bigl[ w_{a \to b} p(n + 1, m - 1) - w_{b \to a} p(n, m) \bigr] \\
        &+ n(m + 1) \bigl[ w_{b \to a} p(n - 1, m + 1) - w_{a \to b} p(n, m) \bigr]\,.
\end{aligned}
\end{equation}
This equation has the same physical reading as
Eq.~(\ref{eq:SatAug30012756PMCEST2025}). Dissipation is also easily
included, e.g., adding a decay rate~$\gamma_1$ to mode~1 adds
$\gamma_1(n+1)p(n+1,m)-\gamma_1np(n,m)$ to
Eq.~(\ref{eq:WedJan7063208PMCET2026}) while a pumping rate~$P_2$ to
mode~2 adds $P_2p(n,m-1)-P_2p(n,m)$, which also have clear physical
meaning.  The steady state of Eq.~(\ref{eq:WedJan7063208PMCET2026})
is found as~\cite{laussy04a}:
\begin{equation} 
  \label{eq:Wed10Aug133031CEST2022} 
  p(n,m)={\xi-1\over\xi^{n+m+1}-1}\xi^n\sum_{k+l=n+m}p(k,l), 
\end{equation}
where~$\xi\equiv w_{b\to a}/w_{a\to b}$ can be related to the
temperature as $\xi=e^{-(E_a-E_b)/k_\mathrm{B}T}$, and is thus $\xi>1$
if~$a$ is the ground state. The sum $P(N)=\sum_{k+l=N}p(k,l)$ on the
rhs of Eq.~(\ref{eq:Wed10Aug133031CEST2022}) is time-independent and
is simply the distribution of the total number of particles in the
whole system. For two thermal states as an initial condition, for
instance,
$P(N)=\sum_{n+m=N}(1-\theta)(1-\nu)\theta^n\nu^m=(1-\theta)(1-\nu)(\theta^{N+1}-\nu^{N+1})/(\theta-\nu)$,
with $\xi=\theta/\nu$ for the equilibrium solution. Those exact
solutions are clearer when reduced to more familiar quantities, such
as the statistics of each state in isolation, which is obtained by
tracing over the other one:
\begin{subequations}
  \label{eq:WedJan21031840PMCET2026}
  \begin{align}
    p_a(n)&=\xi^n(\xi-1)\sum_{N=n}^\infty{P(N)\over\xi^{N+1}-1}\,,\\
    p_b(n)&=\xi^{-n}(\xi-1)\sum_{N=n}^\infty{P(N)\over\xi-\xi^{-N}}\,.
  \end{align}
\end{subequations}
These can be easily computed and plotted, for instance from thermal
states, but whatever distribution one has as initial conditions, one
can see that while the sums in both equations constitute a decreasing
function of~$n$, the ground state distribution $p_a$ has a prefactor
for the sum that is exponentially growing with~$n$ while~$p_b$'s
prefactor is exponentially decreasing with~$n$. This means that the
ground state could feature a peaked distribution where the maximum
probability is not at~$n=0$ (vacuum), unlike the ground state that
exacerbates this characteristic by magnifying the low polariton-number
probabilities, i.e., depleting the state.  This effective
stabilisation of the particle number is the coherence buildup, or
condensate nucleation.  For thermal states, one needs~$\theta+\nu>1$
for $p(0)$ to not be the highest probability, i.e., in terms of
populations, one needs at least two particles on average as a
necessary condition for the ground state to exhibit some
coherence. With an average of 1.9 particles, for instance, one always
finds both states---ground and excited---with vacuum as the most
probable configuration. Besides, as $\xi\to\infty$, $p_a(n)\to P(n)$
and the ground state---a single state---copies the statistics of the
macroscopic system as a whole, making itself a replica not only of the
population but also of its fluctuations: this transfer of the
macroscopic into the microscopic is very much in the spirit of
Einstein's BEC. We have dealt with two modes only, but the mechanism
is general~\cite{laussy12a} and its simplest case clearly shows how
correlating the particle distribution between the states leads to a
stabilisation of their fluctuations, with the possibility for the state
with an influx of particle to develop some coherence when none
preexists in the system, say because we start with all states being
thermal.  With pump and decay, the situation is even more favorable
for condensation and one can now grow an essentially perfect coherent
state in the ground state with only one pumped excited state. This
mechanism of correlating the ground state population-numbers to
excited states can now be coupled to conventional Boltzmann equations,
to study and benchmark the spontaneous coherence buildup of a
realistic polariton laser~\cite{laussy04c}. More interestingly,
Shishkov \emph{et al.}  have carried out the full many-body analysis
of the density matrix of the free lower
polaritons~(\ref{eq:ThuJan1012555PMCET2026}) under phonon dissipative
coupling of the type of Eq.~(\ref{eq:ThuJan8122822AMCET2026}).  The
problem can be solved exactly in the microcanonical ensemble when the
system has exactly~$N$ polaritons, in which case the partition
function for the ideal gas (with parabolic dispersion), say in 2D
(they also work out the 3D case) reads:
\begin{equation} 
  \label{eq:Sat13Aug155139CEST2022} 
  Z_N=\sum_{n=0}^N{1\over n!}B_n\left({1!G_{2D}\over 1^2},{2!G_{2D}\over2^2},\cdots,{n!G_{2D}\over n^2}\right), 
\end{equation}
where~$G_{2D}\equiv mVT/(2\pi\hbar^2)$ is the number of states
above~$\mathbf{k}=\mathbf{0}$ in the energy range of the
temperature~$T$, and $B_n$ is the $n$-th complete exponential Bell
polynomial, which is a way to count weighted set partitions of $n$
elements, rooting the problem deep into combinatorics. From this, they
recover known textbook results such as the fraction of particles in
the ground state, which is
${n_0\over N}\approx 1-{mT\over 2\pi\hbar^2}{V\over N}\ln N$.  The
logarithmic divergence means that the ground-state occupancy is never
macroscopic in infinite 2D systems. In 3D, the $\ln N$ term is
replaced by a constant, namely $\zeta(3/2)$, also with a different
power of the prefactor of $V\ln N/N$, showing that, 3D condensation is
possible. It is textbook knowledge that Einstein's argument for BEC
fails in 2D as the integral converges and all particles can be
accommodated in the available phase space down to zero temperature. In
modern condensed-matter physics, this takes more elaborate
mathematical forms with theorems and/or equalities that forbid
long-range order in two-dimensional
systems~\cite{mermin66a,hohenberg67a}.  Here, we find its resurgence
in a more primitive dynamical theory, which recovers known results
from quite unrelated other approaches, showing that such behaviours
are robust and fundamental. Condensation and coherence remain,
however, possible for finite-size systems~\cite{malpuech03b}, in the
presence of a confining potential~\cite{balili07a} or, in interacting
systems, through the Berezinskii--Kosterlitz--Thouless (BKT)
mechanism~\cite{dagvadorj15a,caputo18a}, binding vortices and
antivortices to establish long-range coherence. All these variations
support polariton condensation in 2D (as well as in lower dimensions).
Another value of Shishkov's theory is to provide quantities not
readily available from other treatments, in particular the linewidth
of the spectrum from the ground state, obtained from the
$\tau$-Fourier transform
of~$\lim_{t\to\infty}\langle\ud{p_0}(t)p_0(t+\tau)\rangle$, which
temporal correlator can be computed from the Lindblad formalism, since
this brings a dynamics of relaxation. In this way, while in 3D they
find that the spectral width follows the Schawlow--Townes relation of
a line narrowing inversely proportional to the population, namely,
$\Gamma_\mathrm{3D}=\Gamma G_{3D}\zeta(3/2)/N$, in 2D, however, they
find a much slower line narrowing
$\Gamma_\mathrm{2D}=\Gamma G_{2D}{\ln N\over N}$ which is a striking
prediction as it is one strictly based on dimensionality, rather than
interactions~\cite{porras03a}, dephasing or other less fundamental
factors that have often been invoked. Their formalism is also ideally
suited to study two-polariton coherence, which was the initial goal of
the formalism that inspired this approach~\cite{laussy04a}, in line
with quantum-optical charactizations of polariton
condensation~\cite{deng02a,adiyatullin15a,klaas18a}. Two-photon
coherence reproduces analytically the simple qualitative tendency of a
transition from a thermal $g^{(2)}=2$ for the ground state (even for a
fixed number~$N$ of polaritons in the entire system) down to the value
of a coherent state as~$g^{(2)}=1+G_\mathrm{2D}/N$. They also find a
``deep quantum regime'' at very small temperature, where coherence is
approached this time from below, $g^{(2)}=1-1/N$, suggesting that all
the particles collapse into the ground state as a Fock state. Since
the normalization of correlation functions---a ratio of
observables---is not obtained as weighted averages in the
$N$-polariton subspaces, the onset of condensation is not concomitant
with the onset of population increase. Therefore, population and
temperature alone are not enough to define the condensate's
coherence. For instance, the type of pumping is critical for
condensation: particles should be pumped into the excited states and
sucked into the ground state to condense, as their direct injection
into the ground state---although more efficient for population
buildup---spoils their coherence. Such fundamental results are
overlooked, if not overshadowed, in more ambitious models which may
reflect them without knowledge that the reported phenomenology might
be rooted in the pure dynamical relaxation of bosons.

\section{Polariton dynamics}

\subsection{Propagation of polariton condensates}
Excitons had already been noted for their unique propagating
properties. At the theoretical level, they emerged as the consequence
that one electronic excitation in a crystal does not stay put to its
site but propagates as a wave into the lattice~\cite{frenkel31a}.
Experimentally, they behave as a gas~\cite{tamor80a} with considerably
higher mobility than other solid-state particles, e.g., drifting under
applied strain over distances of several millimeters at speeds
exceeding several hundred meters per second, as opposed to a few
millimeters per second for an electron. Photons, on the other hand,
can be understood as pure propagation, being pinned to the light
line. Their acquisition of a mass in a cavity according to
Eq.~(\ref{eq:SunAug24023426PMCEST2025}) makes them propagate at much
smaller velocities but still larger than the already very fast
excitons. The propagation of their bound state---the polariton---is
clearly an interesting problem, especially thanks to the ease with
which it can be observed in ultrafast real time and in various
spaces. The propagation of a condensate immediately brings forward the
problem of superfluidity~\cite{kavokin03a}, although BEC and
superfluids are, strictly speaking, independent concepts (one can
occur without the other). There has been a considerable body of work
on the topic~\cite{carusotto04a,lozovik08a,wouters10a,juggins18a}.
Here, we will focus on the most striking---although not most
fundamental---connection between a condensate and superfluidity,
namely, propagation of coherent wavepackets.  The one-dimensional case
is already interesting as capturing many-effects in the simplest
case. Condensates propagating in microcavity wires have been shown to
bounce back off the edges and self-interfere~\cite{wertz12a}. In this
case, the short polariton lifetime was compensated by replenishing
from the exciton reservoir. A more fundamental regime was achieved by
Snoke's group~\cite{steger13a} with their long-lifetime (several
hundred ps) polaritons.  Besides their interest for thermalization of
the condensates, such cavities allow a detailed study of polariton
propagation both below and above threshold, as a function of
real-space, momentum-space, energy and time. This simple problem, even
in the absence of interactions (at low densities), and without the
complications of an exciton-reservoir, is already quite intricate. To
account for the simple time-of-flight from their point of creation
(laser spot) to the distance in space at which polaritons settle into
the~$k=0$ momentum condensate---in a process where they trade their
initial energy and momentum by going uphill a ramp potential (due to
the spatially-varying exciton-photon detuning caused by edges in the
sample wafer)---Steger \emph{et al.}~\cite{steger13a} had to integrate
numerically Hamilton's equations
$\dot x=\partial\mathcal{H}/\partial(\hbar k)$
and~$\hbar\dot k=-\partial\mathcal{H}/\partial x$ for a
spatially-dependent polariton Hamiltonian. In this way, they could
reproduce an observed anomalous slow-down of the condensate as it
escapes from the pumping spot. The model shows that the interplay of
varying effective masses and energy renormalization along the
polariton dispersion, rules the propagation. In fact, even the
simplest free-space propagation of polaritons at constant momentum is
surprisingly rich. This has been studied in detail by Colas and the
author~\cite{colas16a}, with, first, the identification of two
concepts of mass to describe polariton wavepacket dynamics:
\begin{equation}
  \label{eq:WedJan7022416PMCET2026}
  m_1(E,k)\equiv {\hbar^2k\over\partial_k E}\quad\text{and}\quad
  m_2(E,k)\equiv{\hbar^2\over\partial_k^2E}\,,
\end{equation}
where $m_1$ is an \emph{inertial mass} that determines the wavepacket
group velocity~$v=\partial_k E$ from de Broglie's relation $p=\hbar k$
and the classical momentum $p=mv$, while $m_2$ is a \emph{diffusive
  mass} that determines the spreading of a Gaussian wavepacket
according to its standard
deviation~$\sigma(t)=\sqrt{\sigma(0)+(\hbar t/[2m_2\sigma(0)]^2)}$.
While for a massive particle, i.e., a parabolic dispersion, $m_1=m_2$
and one makes no such distinction, the strongly anharmonic dispersion
of polaritons leads to a collection of striking features merely from
the interplay of those two masses, which are shown on the right-hand
side axis of Fig.~\ref{fig:ThuJan1121128PMCET2026}(b) for the lower
polariton. The same could be done with upper polaritons, but lower
polaritons are usually favoured, also for propagation.  The
localisation of polaritons is an ill-defined concept, as only
polariton plane waves can jointly couple a photon plane wave and an
exciton plane wave. Any localisation of one component, say the
exciton, delocalizes the other one to maintain the state on one branch
only, e.g., a $\Psi_\mathrm{ex}(x)=\delta(x-x_0)$ exciton will dress
itself with a sharply peaked but broadened photon field (for which
there is no analytical expression) that spreads around the exciton
point in a fuzzy photonic spot~\cite{colas16a}. The propagation of
this complex composite object with its asymmetric spatial profile is
then ruled by the local dispersion.  At low~$k$, in the parabolic
region of the dispersion, both masses are identical but as one gets
closer to the inflection point, $m_2$ undergoes a first divergence,
and then becomes negative. An infinite diffusive mass means that the
wavepacket does not diffuse anymore and propagates as if a soliton,
without distorting its shape, whatever this might be, since it is
propagating according to a transport equation as opposed to a
self-sustained balance of dispersion and interactions. On the other
side of the inflection point, there is a back-flow and the wavepacket
now interferes with itself, resulting in a self-interfering packet
(SIP) that produces ripples that eventually break into a train of
pulses, each with a different polaritonic quantum state, and with
phase jumps between successive pulses. We thus recognize many
solitonic features in a system which, however, is non-interacting. The
tension between the propagating photon and exciton components can be
so extreme as to cause a breakup in space of the polariton into a
photon and an exciton packet. The Rabi oscillation in time thus gets
projected into space. This dynamics can be better understood by a
wavelet decomposition to vizualize the momentum distribution of the
packet at various spatial points of its propagation. This pinpoints
similarities of polaritons with the well-known family of phase-shaped
self-accelerating wavepackets~\cite{efremidis19a}, making polaritons
dispersion-shaped self-interfering counterparts~\cite{colas20a}. Such
an analysis can also be done in two-dimensions, where it can, when
paired with interactions and anisotropic dispersions, cast a new light
on X-waves~\cite{colas19a}. This picture also allows one to describe
single-branch polariton dynamics as fractional Schr\"odinger
propagation~\cite{colas20b}. While the backflow due to the polariton
dispersion has been evidenced experimentally~\cite{ballarini17a},
there has been no in-depth experimental scrutiny of such basic
effects. They remain important as intrinsic to particles with
indefinite mass~\cite{greenberger74b} and suggest the propagation of
polaritons as fundamental as well as ideal objects to study
low-velocity relativistic effects~\cite{greenberger97a}, in particular
the role of proper time in the global phase dynamics of superpositions
of particles with different masses~\cite{greenberger01a}, which are
nowhere better implemented than with polaritons.






\subsection{Gross-Pitaevskii equation}


The most popular formalism and theoretical approach to tackle the
problem of polariton condensates is that of the Gross-Pitaevskii---or
nonlinear Schr\"odinger---equation (GPE), that has been adapted to
take into account specificities of polaritons.  The simplest extension
to the traditional atomic case reads:
\begin{multline}
  \label{eq:TueJan6121828PMCET2026}
  i\hbar\frac{\partial\Psi(\mathbf{r}, t)}{\partial t} 
  = \left(-\frac{\hbar^2\nabla^2}{2m} + V(\mathbf{r}) 
  + g |\Psi(\mathbf{r}, t)|^2 \right)\Psi(\mathbf{r}, t) \\
  -i\frac{\hbar\gamma}{2}\Psi(\mathbf{r}, t)
  + F(\mathbf{r},t)e^{-i\omega t + \mathbf{k}\cdot\mathbf{r}}
\end{multline}
which we have split to separate, in the first line, the atomic GPE---in
the presence of a potential~$V$ for confinement---from, in the second line,
the driven-dissipative aspect of polaritons, namely, the decay (at
rate~$\gamma$) and pumping (inhomogeneous term).
This phenomenology is sufficient to already capture a large breadth of
polariton condensate phenomenology under strong pumping, including
superfluidity~\cite{amo09a,amo09b},
vortices~\cite{nardin11a,sanvitto10a,sanvitto11a} and a huge body of
solitonic behaviours of all types~\cite{amo11a,sich12a}. Written in
polar form, with the so-called Madelung
decomposition~$\Psi(\mathbf{r}, t) = \sqrt{N(\mathbf{r}, t)} e^{ i
  \Theta(\mathbf{r}, t)/\hbar}$, it also brings polariton condensates
into the so-called Kardar-Parisi-Zhang (KPZ) universality class, by
mapping the dynamics of the phase at long distances to the eponymous
equation for stochastic interface growth~\cite{deligiannis20a}. Three
regimes have been identified~\cite{helluin25a}: i) the
Edwards-Wilkinson regime of weak nonlinearities, with large phase
fluctuations, ii) the KPZ regime, where strong nonlinearities lead to
a superdiffusive growth of the fluctuations and roughening of the
phase~\cite{fontaine22a}, and iii) BKT with both phase and density
fluctuations, leading to proliferation of vortices~\cite{caputo18a}.
Polaritons proved instrumental in demonstrating KPZ universality in
2D~\cite{widmann26a}. In GPE treatments, the dispersion is often
assumed to remain parabolic, although the exciton-photon structure has
also been considered, for instance, in the early works by Carusotto
and Ciuti~\cite{carusotto04a} or Whittaker~\cite{whittaker05b} to
describe the spectra of elementary excitations of optical parametric
processes.  Striking phenomenology going beyond the single-field GPE
includes Voronova and Lozovik~\cite{voronova12a}'s description of
largely differing exciton and photon healing lengths from the same
polariton condensate, with two-order of magnitude smaller excitonic
healing lengths resulting in excitons hiding inside the photon vortex
core, or Komineas \emph{et al.}~\cite{komineas15a} pointing out a new
phenomenology of discontinuous band of dark solitons, where the
exciton field exhibits a jump at the soliton center.  There are many
other works in a similar vein~\cite{kartashov12a,elistratov16a}, for
instance considering a single field but with non-parabolic
dispersion~\cite{pinsker15a}. This rich phenomenology suggests that
the exciton-photon structure of polaritons in GPE physics has garnered
insufficient experimental interest.  Another departure from the atomic
case even in its one component version, comes with the inclusion of
the exciton reservoir, which often plays an important role, while the
ground state, if condensed, can still be described using the GPE. This
has been combined as follows:
\begin{subequations}
  \label{eq:TueJan6022248PMCET2026}
  \begin{align}
    i\hbar\frac{\partial\Psi(\mathbf{r}, t)}{\partial t} 
    &= \Biggl(-\frac{\hbar^2\nabla^2}{2m} + V(\mathbf{r}) + g |\Psi(\mathbf{r}, t)|^2 \Biggr) \Psi(\mathbf{r}, t) \notag \\
    &\quad + i\frac{\hbar}{2}\bigl[R n_R(\mathbf{r},t) - \gamma_p\bigr] \Psi(\mathbf{r}, t),
    \label{eq:GP-like} \\[1.2ex]
    \frac{\partial n_R}{\partial t}(\mathbf{r},t) 
    &= -\frac{n_R(\mathbf{r},t)}{\tau_R} - R n_R(\mathbf{r},t)|\Psi(\mathbf{r}, t)|^2 + P(\mathbf{r},t)\,.
    \label{eq:reservoir}
  \end{align}
\end{subequations}
where we still recognize the standard (atomic) GPE at the top line,
which now gets upgraded with an incoherent, and positive, contribution
describing pumping, in effect counter-balancing decay. While the
latter is simply a constant, the reservoir pumping is both
space~$\mathbf{r}$ and time~$t$ dependent, being directly proportional
to the population~$n_R(\mathbf{r},t)$ of excitons available to scatter
into the condensate. In turn, one needs an equation for this
reservoir, provided by Eq.~(\ref{eq:reservoir}) in a physically
transparent form, featuring in particular Bose stimulation of exciton
into the condensate. This spatial dependence results in a self-carving
of a potential by the condensate itself, thanks to the much larger
exciton mass which prevents their diffusion and allows them to follow
dynamically the condensate's evolution. The reservoir can also be
created optically. This allows for engineering potentials with complex
spatial profiles.  The density-dependent blueshift of the energy
landscape enables self-trapping: condensed polaritons modify and
deepen their own confining potential around the pump spots,
stabilizing themselves and forming bound and phase-locked states
between the injection sites~\cite{tosi12a}. Using ring-shaped
reservoir potential geometries, it is also possible to decouple the
polariton condensate from the exciton
reservoir~\cite{askitopoulos13a}.  In this way, polaritons in
extremely high-quality samples could form condensates of world-record
spatial extension, shielded from the exciton
reservoir~\cite{ballarini17a}.  There are essentially two approaches
to tackle the Gross-Pitaevskii equation. One is to linearize it and
study the stability of its spectrum of elementary
excitation~\cite{wouters07c}. The other is to solve the full equation
numerically~\cite{voronych17a,wingenbach25a}. Often those are brought
together to solve the nonlinear problem, find its steady states and
determine their dynamical stability.  Other models combine the GPE
with incoherent relaxation of a different type than the previous
reservoir model. Saltykova \emph{et al.} for instance introduced pure
relaxation terms in the form of a gradient to the phase part~$\Theta$
of the Madelung representation, which describes the important phonon
relaxation~\cite{saltykova25a}. This problem had been tackled by
Solnyshkov \emph{et al.}~\cite{solnyshkov14a} in a hybrid
Boltzmann--GPE model, which, however, violates particle-number
conservation expected from a pure-energy relaxation term. Saltykova's
modified GPE equation simply adds to the top line of
Eq.~(\ref{eq:TueJan6121828PMCET2026}) a
term~${}+{i\hbar\over2}\lambda\Psi[\Psi^*\nabla^2\Psi-\Psi\nabla^2\Psi^*]$
(it can also keep the other terms to describe those other
processes). Those theoretical models have not yet been put under the
scrutiny of experimental investigations, although they typically
result in strong qualitative predictions, say on the stability of
superfluidity, localisation of propagating packets or slowing down of
their speed. There are other theoretical models aiming to generalize
even more the GPE. For instance, a popular formalism to go beyond
mean-field and include some dynamics of the quantum fluctuations is
the truncated Wigner approximation, introduced for polaritons by
Wouters and Savona~\cite{wouters09a}, which averages over various
stochastic trajectories and allows to access quantum observables such
as two-photon correlations.

\subsection{Josephson oscillations}

The physics of two coupled condensates (or coherent states) in the
presence of a Kerr nonlinearity, can be described as a type of
Josephson dynamics, which initially involved Cooper pairs from two
superconductors in contact, but can also be formulated for bosonic
Josephson Junctions~\cite{gati07a}.  Like unconventional blockade,
which can be implemented in a variety of platforms, the Josephson
effect for bosons can be realized in a plethora of forms, from
spatially separated condensates~\cite{wouters08b} to internal
oscillations~\cite{voronova15a}, as is the case for photons and
excitons within the polariton.  This is a precursor of the phenomenon
of superfluidity as the dynamics between the two modes is driven by
the relative phase between the two condensates, and the two best
observables are indeed the relative phase~$\theta$ and the population
imbalance~$\rho$ between them.  The clear-cut terminology of
superconducting Josephson oscillations can be adapted to some extent
to bosons, in which case the so-called ac Josephson effect---or
oscillations of current under applied voltage---occurs in the presence
of detuning of the bosonic modes, while so-called plasma
oscillations---intrinsic instabilities of the junction---occur due to
the initial condition not being a stationary solution of the coupled
condensates.  This is a fairly complex terminology to describe a
ubiquitous phenomenology of coupled oscillators: oscillations.  Three
regimes have been identified: i) non-interacting (Rabi), ii) weakly
interacting (Josephson) and iii) strongly interacting (Fock).  The
behaviour of the phase---running (with discontinuity if wrapped
modulo~$2\pi$) or oscillating---has been regarded as a criterion to
set apart the Rabi~\cite{lagoudakis10a} from the
Josephson~\cite{abbarchi13a} regimes, both of which have been observed
with polaritons, following the atomic precedents. Recently, picking up
on concerns from the atomic literature, Voronova \emph{et al.}
challenged that Josephson junctions can be defined for merely two
localized condensates in contact, and argued that such a terminology
should be used only for macroscopic systems (of infinite size so that
oscillation frequencies become system independent), or if the barrier
between them can cancel completely their tunneling, like an insulator
breaks the superconducting phase in the original effect. This results
in a phase jump as the density cancels
exactly~\cite{voronova25a}. When this is not the case, the phase
varies smoothly inside the barrier (a weak-link) and Voronova regards
this dynamics as that of a superfluid, which she calls the
``hydrodynamics regime''.  The vanishing density and phase-jump can
also occur in Josephson oscillations between two modes only, either in
the pure, full-amplitude Rabi oscillations, or more generally, when
they are dissipative. In both cases, if one condensate gets to deplete
completely, the phase becomes ill-defined and the phase jumps. In the
dissipative case, it also typically switches between a running and an
oscillating phase. All this phenomenology can be unified and clarified
by tracking the dynamics, not of each condensate, but of the effective
oscillator constituted by their relative phase and relative
population. Furthermore, the representation should not be of one
variable as a function of the other, but as a trajectory on the
canonical geometry for a two-level system: the Bloch sphere, as shown
in Fig.~\ref{fig:ThuJan1121128PMCET2026}(c). The equation for the
Josephson dynamics then simply reduces to~\cite{rahmani16a}:
\begin{equation}
  \label{eq:TueSep23080643PMCEST2025}
  |\langle\ud{p_\theta}q_\theta\rangle|^2+\rho_\theta^2=(N/2)^2
\end{equation}
where $N$ is the total
population~$\langle\ud{p_\theta}p_\theta\rangle+\langle\ud{q_\theta}q_\theta\rangle$
for the polariton operators~$p_\theta$ and~$q_\theta$ previously
denoted as diagonalizing the non-interacting Rabi Hamiltonian, i.e.,
they are the polaritons for a given detuning, with the mixing angle
now made explicit. They define two poles on the Bloch sphere, which in
turn define a vector~$\vv*{\rho}{\theta}$. The initial condition---or
polaritonic quantum state---defines the distance~$\rho_\theta$ along
this vector from the center of the sphere. This is $\pm1$ when in a
polariton state (for any given detuning) with the trajectory reducing
to a point: polaritons are stationary states. While this connection
is never made, this would correspond to the dc Josephson effect, of a
continuous current driven by a fixed phase difference, although the
phase here is~$0$ or~$\pi$ and both bare states (exciton and photon
condensates in the internal Josephson, or the two spatially separated
condensates), stabilize their population by a simultaneous
bidirectional flow of bare excitations. Any other configuration
results in oscillations, which can be strongly anharmonic although
Eq.~(\ref{eq:TueSep23080643PMCEST2025}) is, in absence of
interactions, the equation for a circle on the Bloch sphere. Now comes
the link between this trivial dynamics, and the one ``observed''
through a laboratory observable, e.g., the photon
component~$\langle a\rangle$ for internal oscillations or the order
parameter for each of the condensates~$\langle a\rangle$
and~$\langle b\rangle$ for spatial junctions. Such observables are
retrieved by projection on the laboratory axis~$\vv\rho$ which
intersects the bare states on the Bloch sphere. This explains
anharmonicities of oscillations for an otherwise harmonic motion on
the sphere. The running or oscillating phase also follows from this
topology, namely, whether the circle on the sphere encircles (running
phase), or not (oscillatory), the laboratory axis~$\vv\rho$. This
makes it clear that such a phenomenology is therefore a
basis-dependent property rather than a criterion for Rabi or Josephson
regime. The vanishing population of one condensate is similarly an
artifact of the projection when the circle happens to touch the
laboratory axis. Changing the frame of reference makes such dramatic
cancellations of populations and phase disruptions no longer
occur. The intrinsic, canonical dynamics remains, throughout, a simple
circle on the sphere. Now bringing in interactions, the trajectory
distorts from a circle into a warped trajectory, still on the
sphere. Similarly, whether this trajectory loops around or not the
observable axis results in a running phase, or not. Clearly, this
criterion does not give a sound definition for the Rabi and Josephson
regimes. Instead, Rahmani and the author proposed~\cite{rahmani16a} to
distinguish these two regimes based on the number of fixed points of
the dynamics: two (the polaritons) for the Rabi regime and four for
the Josephson regime. This is unambiguous and well-defined on the
sphere, i.e., regardless of any choice of basis for
observation. Polaritons are centers, i.e., complex-valued eigenvalues
with trajectories oscillating harmonically without attraction or
repulsion. In the Josephson regime, one of the additional points is a
saddle point (the other three remain centers), that is, whose
eigenvalues with real parts of both signs result in both attraction or
repulsion of the trajectories. The remaining important effect,
self-trapping---where oscillations quenched by the interactions
localize the trajectory around one condensate---is thus observed as a
result of this instability sending back the trajectory to its
center. Whether the dynamics of the effective oscillator formed on the
two sides of Voronova's weak link also fits this picture, has not been
addressed. This bridges superfluidity between its simplest (two modes)
case and a more traditional flow in real, continuous space.

\section{Outlook}

While we have tried to capture a bit of the most recent breakthroughs
relying on polariton condensates, we had to leave aside a considerable
amount of important and exciting physics that most recently
emerged---such as bound state in the continuum~\cite{ardizzone22a},
turbulence~\cite{panico23a}, time crystals~\cite{carrarohaddad24a} or
supersolids~\cite{trypogeorgos25a}--but some of which can be described
briefly in some final outlook. Our most serious offense is certainly
to have overlooked the more literal sense in which the polariton is a
spinor field: through its spin.  There is a rich spin dynamics due to
strong spin-anisotropic interactions (due to the exciton exchange
term), and the sensitivity to the cavity's TE-TM mode splitting, which
provides a momentum-dependent effective magnetic field, investigated
in-depth in particular by Kavokin and collaborators over the
years~\cite{kavokin04a,shelykh05a,shelykh10b,sedov19a,sedov21a}.  Some
highlights include the Optical Spin Hall Effect~\cite{kavokin05b},
which implements a spin-dependent polariton propagation, causing
polarization rotation and patterns that have been experimentally
confirmed~\cite{leyder07a}, or a rich physics of spinor condensates
which, when combined with the hallmark of
superfluiditity---vortices---yield half-vortices, proposed by
Rubo~\cite{rubo07a} and also confirmed
experimentally~\cite{lagoudakis09a,dominici15b}.  A notable incursion
of polariton physics into unanticipated domains---albeit still in the
wake of pioneering atomic cases---concerns black holes, bringing
analogue gravity with a quantum fluid to polariton condensates.  These
analogs create ``acoustic horizons'' where the polariton flow
transitions from subsonic to supersonic, mimicking event horizons and
enabling Hawking radiation and other quantum gravity effects in a
controlled optical table environment.  The concept originated
theoretically for polaritons with Solnyshkov \emph{et
  al.}~\cite{solnyshkov11a} who also proposed wormholes in spinor
polariton condensates, and shortly after with Gerace and
Carusotto~\cite{gerace12a} in 1D, all treating fluctuations as
Bogoliubov excitations on top of a mean-field background described by
a GPE.  Such fascinating theoretical proposals have been followed by
experimental
implementations~\cite{nguyen15a,jacquet20a,jacquet22a,falque25a},
which in turn rekindled theoretical interest~\cite{svetlichnyi24a}.
Current investigations consider stimulated and amplified Hawking
radiation, or---bringing vortices in this regime---the realization of
so-called extremal (maximally spinning) and even super-extremal
(overspinning) polariton black holes. This would allow one to expose
the spacetime singularity (removing the event horizon), violating
Penrose's cosmic censorship conjecture and thus pushing this already
extreme field of physics way beyond its traditional limits.  We also
overlooked classical devices, despite impressive results often setting
new world records, such as the all-optical polariton
transistor~\cite{ballarini13a} whereby a weak gate polariton beam can
control the flow of a stronger source-drain polariton fluid via
stimulated scattering and nonlinear interactions, achieving optical
gain up to 19 times amplification. A breakthrough was the
demonstration of cascadability: the output of one transistor can drive
the input of another one, a crucial requirement for integrated logic
circuits. Combining multiple inputs, universal logic gates have thus
been demonstrated.  Another landmark is a sub-femtojoule electrical
spin-switch~\cite{dreismann16a} allowing an electric field to control
the condensate's spin, switching the emitted light's polarization with
energies below 0.5fJ per switch, which is several orders of magnitude
lower than conventional spintronic devices. While those constitute
fantastic achievements from polariton condensates, they remain
technical prowess of an engineering character, so we do not enter into
further details. Instead, similar devices but that border on, or even
claim, quantum functionalities, are more noteworthy, also because they
are prone to controversy.  Although we discussed at length the problem
of the quantumness of polaritons, we have not covered works oriented
on applications, as natural consequences of a positive resolve of this
assumption. Would polaritons operate in a genuine quantum regime (of
nonlocal entanglement), they would become the most apt platform for
applications. The fact that relatively modest attention is granted
them from that perspective, is an indication of the little credit they
are still given regarding such quantum features. This will probably
change as technology matures.  Early discussions came from
Ciuti~\cite{ciuti04a} and Savasta \emph{et al.}~\cite{savasta05a} on
entanglement from which-way indeterminacy in the emission of polariton
pairs. Like for condensation, polariton interaction has been hailed as
the key element to power quantum polaritonics, which are
systematically described as strongly-interacting particles, which is
true with respect to photons (which do not interact at all) but not
with respect to ions, atoms, etc. In fact, such weak interactions
might be an asset rather than a curse. Just like polariton blockade
actually requires weak interactions to maintain good squeezing and
optimize the squeezed/coherent interferences, a scheme of quantum
polaritonic interferometry~\cite{nigro22a} was recently shown to
provide high-fidelity deterministic quantum gates~\cite{scala24a}
which degrade with increasing interactions. This suggests a
generalization of the polariton blockade scenario but implementing a
logical truth table in dual-rail logical encoding as opposed to merely
suppressing a two-photon component of the wavefunction. If confirmed,
this would make polaritons most suited for quantum computation in the
regime where they actually are the best: as weakly-interacting
strongly-interfering (WISI) particles. Still, the incentive to
maximize interactions in a more traditional fashion of a single
particle controlling another remains strong and progressing.
Single-polariton operations, either involving Fock
states~\cite{cuevas18a} or merely on
average~\cite{zasedatelev21a,kuriakose22a}, are now becoming
commonplace. Beyond polaritons qubits---which we have already
cautioned should currently still be understood as cebits---quantum
information processing with polariton condensates falls within the
continuous-variable framework. One can mention for illustration Byrnes
\emph{et al.}~\cite{byrnes12a}'s proposal to encode the
qubit~$\alpha\ket{0}+\beta\ket{1}$ as a ``BEC qubit''
$|\alpha,\beta\rangle\!\rangle\equiv(\alpha\ud{a}+\beta\ud{b})^N\ket{0}/\sqrt{N!}$
(although one is dealing here with a Fock state rather than a
condensate) with a speed-up of gate operation by~$N$ (the number of
particles) in a scheme that allows universal quantum
algorithms. Although the authors have polarization in mind for
polaritons, the exciton-photon ``polaritonic'' quantum state appears
particularly relevant. Ghosh and Liew~\cite{ghosh20a} considered
instead the quantum fluctuations on top of the mean field (coherent
state) with population~$N_c\gg1$, with qubit states~$\ket{0}$
and~$\ket{1}$ mapped to the two lowest excited states of each
condensate (one condensate per qubit), for which the effective
Hamiltonian is obtained as
$H_q \equiv \mathbf{E} \cdot \boldsymbol{\hat{\sigma}}$
where~$\mathbf{E}\equiv\big(P\sqrt{N_c}\cos(\varphi),P\sqrt{N_c}\sin(\varphi),(U_a-\omega)/4\big)$
and $\boldsymbol{\hat{\sigma}}$ is the vector of Pauli matrices in the
fluctuation-qubit basis, with also~$\Omega_a\equiv Pe^{i\varphi}$
and~$\omega$ some effective detuning of the driving laser, including
blueshift. This shows that by controlling the pulse intensity, phase
and duration, one can perform, in principle qubit rotation of the
fluctuations.  The higher the coherent state population, the better
the fidelity of the quantum bit operations on its fluctuations.
Coupling two neighboring pillars by tunneling, two-qubit gates can be
similarly implemented. While it is not entirely clear how the qubit
readout of such, supposedly virtual, fluctuations is effected, the
authors still envision polariton-qubit lattices of $100\times100=10^4$
qubits, each operated at a fraction of a mW resulting in a prospective
polariton-condensates quantum computer in a mm$^2$ sized sample
operating at high temperature with less than a kW power, outperforming
in all respects superconducting qubit counterparts.  There is also a
sustained activity on polariton simulators---with a tacit although
seldom justified ``quantum'' aspect---exploiting their facility for
designing potential landscape which can encode a variety of 1D and 2D
Hamiltonians by trapping condensates in various sites and letting them
interact by controlled hopping through tunneling.  Highlights include
Berloff \emph{et al.}'s implementation of a XY
simulator~\cite{berloff17a}, relevant for NP-hard optimization
problems and phenomena like frustration, spin liquids, or the BKT
transition. The successful demonstration of various XY phases on a
lattice of polariton condensates with 45 sites which appears to be
reasonably scalable, supports the relevance of polariton condensates
to provide advantage of some sort, whether analogue, partially quantum
or involving full-fledged quantum supremacy, especially within
tailored schemes such as gain-dissipative
algorithms~\cite{kalinin18a}.  The confinement allowed by etching
cavities to realize a variety of lattices with great flexibility for
band-engineering, also makes polariton condensates at the forefront of
topological photonics, say in honeycomb lattices of microcavity
pillars (so-called ``polariton graphene''), giving rise to
unidirectional, topologically protected edge states that propagate in
one direction only~\cite{nalitov15a}.  The ease of heterostructure
growth allows the design of aperiodic potentials, such as a Fibonacci
sequence~\cite{tanese14a}, in which case a fractal energy spectrum is
observed, with the opening of minigaps that obey the gap-labeling
theorem (a fundamental result of the spectral theory of Schrödinger
operators regarding the density of states in aperiodic systems) or
regular but peculiar geometries such as Lieb lattices. Those feature
three sites per unit cell---depleting a square lattice to position one
corner facing two edge centers---that also yield distinctive band
spectra, including Dirac cones (linear dispersion) intersected by a
flat band (zero dispersion) with infinite mass.  The resulting
vanishing kinetic energy in flat bands gives way to interactions and
disorder.  The straightforward optical characterization of the various
resulting phases, including condensation~\cite{baboux16a}, is another
implementation of polariton simulation.  Almost by definition,
polaritons are also a suitable platform for the physics of
Non-Equilibrium Steady States (NESS), where macroscopic properties
remain time invariant, but as a result of underlying energy/momentum
fluxes that keep the system away from thermodynamic equilibrium, where
steadiness is instead maintained by the principle of detailed
balance. This is an important scientific concept since it is central
to complex systems, in particular biological ones, like living
cells~\cite{bianca12a}, reconnecting to Fr\"ohlich's seminal dynamical
BEC. Polaritons might be the simplest, and thus most fundamental,
example of NESS systems, implementing for instance basic thermal
machines~\cite{toledotude24a}. In this framework, polariton
condensation brings forward remarkable departures from atoms, e.g.,
the possibility to accommodate multiple macroscopically coherent
stable states or sustain various types of relaxation
oscillations~\cite{degiorgi14a,lukoshkin23a}.  In fact, polaritons
might be closer to biological dynamics than to their atomic
counterparts, and they thus could come to play an increasingly major
role at the frontier of quantum physics, complex systems and
biophysics.  Finally, concluding with our observation that polaritons
had finally caught up with Hopfield's neural networks, we must mention
the timely and promising activities in those areas, starting with the
pioneering insight of Liew \emph{et al.} of a polariton neuron, as
early as 2008~\cite{liew08a}, well over a decade before Artificial
Intelligence would spread to all fields of science. The idea then was
to create guided channels (``neurons'') for the propagation of
polarized polariton signals, based on their polarization
multistability. An extension to this scheme that remained heavily
rooted in classical logic was later put forward, also by Liew, this
time with Espinosa-Ortega~\cite{espinosaortega15a}, and with a design
for a full perceptron with trainable weights, inaugurating polaritonic
machine learning through physical Hebbian plasticity.  Polaritons have
subsequently joined the unanimous embrace of AI, where they bring
ultrafast, low-energy and highly-nonlinear optical processing,
possibly at room temperature, with demonstrated applications for
instance in Polariton-based Neuromorphic Networks. Those involve
polariton lattices of condensates which provide the neural network
hardware, powered by ballistic propagation of condensates with
unmatched performances in speed (picosecond timescales) and energy
efficiency (sub-pJ). The theoretical foundations have been established
by Opala \emph{et al.}~\cite{opala19a} with a compelling demonstration
by Ballarini \emph{et al.}~\cite{ballarini20a}'s polaritonic
neuromorphic computing. A nonlinear network of polariton condensates
at sites etched into the cavity can implement reservoir computing---a
type of recurrent neural network where the collection of nodes is
predetermined and the network functions are determined by a single
layer of weights applied to the output~\cite{matuszewski21a}. This is
known to be efficient for temporal or pattern recognition tasks, and
the polaritonic implementation was indeed shown to outperform
traditional linear classifiers on the problem of recognizing
handwritten
digits. 
Those are only examples of resounding successes of polaritons in
emerging technologies, where they achieve no less than world-record
performances.

\section{Conclusions}

We have provided a personal overview of several trends of polariton
condensates, with a focus on i) what makes them special within the
broader field of Bose-Einstein condensation of closed systems at
equilibrium, and ii) emphasizing departures within the field
itself. This highlights linear effects and the need for more careful
and accurate treatment of the quantum nature of polaritons.  The
polariton community tends to align itself more with the atomic
paradigm than it should, putting more importance on interactions than
always needs to be the case, and less on the quantum-optical side,
which is more prominently determined by linear effects such as
interferences, correlations or the dynamics of particles with
anharmonic dispersions, as a result of their momentum-varying
mass. This does not tamper that strongly-correlated quantum many-body
polaritonics remains an apex of the field. While we tried to cover a
broad breadth of representative topics, we strived to delve into the
theoretical arguments and their conceptual
interpretation. Consequently, we left almost untouched large swathes
of the polariton condensates physics, both popular ones, such as
hydrodynamics, solitons, and more offbeat topics, such as weak
lasing~\cite{aleiner12a,zhang15a}, bosonic
cascades~\cite{liew13b,liew16a}, polariton
interferometers~\cite{nigro22a,baryshev25a},~etc.

\bibliographystyle{naturemag}
\bibliography{sci,Books,arXiv}

\end{document}